\documentclass[onecolumn,amsmath,amssymb,11pt,superscriptaddress,nofootinbib]{article}
\usepackage[a4paper, total={6in, 9in}]{geometry}
\usepackage[utf8]{inputenc}
\usepackage[english]{babel}

\usepackage[nottoc]{tocbibind}
\usepackage[inline]{enumitem}

\usepackage[square,numbers]{natbib}
\usepackage{longtable, mathtools}
\usepackage{amsmath}
\usepackage{amsfonts}
\usepackage{amssymb}
\usepackage{makeidx}
\usepackage{stackrel}

\usepackage[]{graphicx}
\usepackage{subcaption}

\usepackage{rotating,booktabs}

\usepackage{color}
\usepackage{lmodern}
\usepackage{kpfonts}
\usepackage{xfrac}
\usepackage{array}
\usepackage{url}

\usepackage{amsthm}
\newtheorem{theorem}{Theorem}
\newtheorem{corollary}{Corollary}

\usepackage{setspace,caption}
\usepackage{lipsum}

\usepackage{lineno} 

\usepackage{mwe}    
\usepackage{float}
\usepackage[linesnumbered, ruled]{algorithm2e}
\SetKwRepeat{Do}{do}{while}%
\usepackage{algpseudocode}
\usepackage{comment}
\IncMargin{1em}
\usepackage{mwe}

\usepackage{xr}
\usepackage{bbm}

\title{Stochastics of DNA Quantification}
\author{ Abdoelnaser M Degoot and Wilfred Ndifon\footnote{Address for correspondence: wndifon@aims.ac.za}\\
African Institute for Mathematical Sciences, Next Einstein Initiative, Rwanda }
\date{\today}

\begin{document}
	\maketitle
	
	\section{Abstract}
	A common approach  to quantifying DNA involves repeated cycles of DNA amplification. This approach, employed by the polymerase chain reaction (PCR), produces outputs that are corrupted by amplification noise, making it challenging to accurately back-calculate the amount of input DNA. Standard mathematical solutions to this back-calculation problem do not take adequate account of such noise and are error-prone. Here, we develop a parsimonious mathematical model of the stochastic mapping of input DNA onto experimental outputs that accounts, in a natural way, for amplification noise. We use the model to derive the probability density of the quantification cycle, a frequently reported experimental output, which can be fit to data to estimate input DNA. Strikingly, the model predicts that a sample with only one input DNA molecule has a $<$4\% chance of testing positive, which is $>$25-fold lower than assumed by a standard method of interpreting PCR data. We provide formulae for calculating both the limit of detection and the limit of quantification, two important operating characteristics of DNA quantification methods that are frequently assessed by using \textit{ad-hoc} mathematical techniques. Our results provide a mathematical foundation for the rigorous analysis of DNA quantification.	
	
	\section{Introduction}
	The quantification of genomic targets is of interest in a large variety of applications in biology, biotechnology and medicine, from determining an individual's disease status to detecting minute changes in gene expression profiles occurring across space and time (eg. \citep{wolfel, wang1989}). This is typically achieved by converting non-DNA genomic targets into DNA, which is then amplified to enable its quantification. In principle, this allows even small numbers of genomic targets to be accurately measured. However, in practice, the DNA amplification process, being stochastic, generates outputs that contain noise. Accurate measurement, therefore, requires an adequate, quantitative understanding of this noise. Thus far, this has proved challenging to achieve.
	
	A specific and very popular instance of a DNA quantification method is the real-time polymerase chain reaction (PCR)  \citep{saiki, higuchi}. In PCR, DNA molecules are repeatedly amplified in a cyclic manner. As they are amplified, fluorescently labeled nucleotides are incorporated into the newly formed DNA molecules, increasing the overall fluorescence emitted. The resulting fluorescence profile is used to determine the quantification cycle (denoted $Cq$ or $Ct$ value), at which the number of molecules exceeds a defined threshold, called the quantification threshold. A PCR reaction is considered to be positive if its $Ct$ value is less than or equal to the maximum possible cycle. Despite the fact that the $Ct$ value is only an indirect readout of the number of input DNA molecules, it is often the only reported output of PCR experiments. A variant of conventional PCR, called digital PCR \citep{dpcr0}, uses the fraction of positive reactions to estimate the number of input DNA molecules. To this end, it assumes that a reaction is positive if and only if it contains at least one target molecule. It is unclear under what conditions this assumption is valid, and when it must be discarded in favor of a more realistic alternative.
	
	Here we describe a parsimonious mathematical model that is useful for analysing the DNA quantification process, and for guiding the interpretation of experimental outputs. We use PCR as an example, although our analysis is applicable to other methods such as loop-mediated isothermal amplification of DNA \citep{lamp}. Experiments indicate that the PCR process exhibits different phases, characterized by different efficiencies of DNA amplification. Therefore, we construct a mathematical model of a PCR process with an arbitrary number of phases, each with its own amplification efficiency. We use this model to obtain the following results:
	\begin{itemize}
			\item We derive the generating function for the probability distribution of the number of molecules found in a PCR experiment at an arbitrary time $t$. We also derive the probability density function (pdf), mean, variance, and cumulative density function (cdf) of the $Ct$ values produced by such an experiment. Either the pdf or the cdf can be fit to PCR data to estimate the number of input DNA molecules.
			\item In the simplest instance of our model -- a single-phase PCR model that accounts for amplification noise but not for (upstream) DNA sampling noise -- the mean $Ct$ value, given by $(\psi(x+1) -\psi(n))/r$, is well approximated by $\ln(x/n)/r$ \citep{mutesa} when $n \gg 1$, where $n$ is the number of input molecules, $r$ (defined on a base-$e$ scale) is the amplification efficiency, $x$ is the quantification threshold, and $\psi(\cdot)$ denotes the digamma function.
		  \item We provide a formula for calculating the limit of detection (LoD) of a PCR experiment, that is, the smallest number of input molecules that can be detected with a failure rate not exceeding $\alpha$. Using a single-phase PCR model, we find that when $\alpha = 0.05$, the LoD increases from 3, the value determined while accounting for sampling noise only, to $\approx$10 when both sampling noise and amplification noise (with $r$ set to 95\% of the maximum possible efficiency, m.p.e.) are considered. The LoD increases as $r$ decreases, doubling to $\approx$20 at 90\%  m.p.e. This illustrates the under-appreciated, dramatic effect that amplification efficiency has on the LoD. 
			\item We provide a formula for calculating the limit of quantification (LoQ) of a PCR experiment, that is, the smallest number of molecules that can be quantified with a defined level of precision and a given maximum failure rate $\alpha$. Counter-intuitively, the single-phase PCR model predicts that the LoQ does not depend on amplification efficiency. When $\alpha = 0.05$, the LoQ increases from $10$, obtained when up to a two-fold deviation from the expected number of input molecules is allowed, to 820, when at most a 10\% deviation is allowed. This indicates that 10 or fewer molecules cannot be measured with a better than 2-fold error more than 95\% of the time.
			\item The model indicates that a key assumption commonly used when interpreting digital PCR data -- that a PCR experiment with only one input molecule will always produce a positive outcome -- is  invalid under a wide range of conditions. Even when the amplification efficiency is set to a high value of 95\% m.p.e, the probability that such an experiment will yield a positive outcome is predicted to be $<$4\%. We describe two different approaches by which accurate estimates of the number of input DNA molecules may be obtained from digital PCR data.
			\end{itemize}
			
		It should be noted that there have been previous attempts to improve the interpretation of PCR data through mathematical modeling. The classical approach to estimating the amount of DNA found in a focal sample involves comparing data generated by that sample versus data obtained from a reference sample containing either a known or an unknown amount of DNA \citep{livak}. The need for a reference sample with a known amount of DNA, the determination of which is itself subject to experimental error, makes accurate absolute quantification of DNA found in the focal sample challenging. An alternative approach involves fitting mathematical models, mostly phenomenological in their construction, to PCR data generated by the focal sample alone \citep{higuchi, livak, methods2, LRE, sigmoid4}. See \citep{comparison1} for a comparison of various methods based on this approach. None of these methods provides an adequate accounting of how amplification noise shapes PCR data.
		
		The remainder of this paper is organized as follows: We provide an overview of the model's structure in Section \ref{preliminaries} and present our main mathematical results in Sections \ref{n_known} and \ref{lam_known}. We apply these results to compute the LoD and LoQ in Section \ref{lodq}, and we investigate how amplification noise complicates the accurate interpretation of digital PCR data in Section \ref{dpcr}. We summarize the results and discuss other applications of our methods in Section \ref{discuss}. To improve readability, we only present mathematical proofs and detailed calculations in the Appendix (Section \ref{appendix}).
	
	\section{Results} \label{res}
	\subsection{Preliminaries}\label{preliminaries}
		We model the PCR process as a continuous-time, discrete-state Markov jump process \citep{ckm} evolving up to time $T$. This representation of the PCR process is based on the facts that (1) the primary products of PCR reactions, DNA molecules, are countable, and (2) what happens in the next cycle of the reaction is conditionally independent of what happened in the past given the present state of the reaction. Our decision to make time continuous (rather than discrete) is based on the fact that experimentally measured $Ct$ values are positive real numbers. As a consequence, reaction rates are defined in base $e$ instead of base $2$ (expected for a discrete-time PCR process), but it is straightforward to convert between these two bases.
  
        We divide the time interval $[0,T]$ of the PCR process into $p$ non-overlapping subintervals $I_i$, each one corresponding to a distinct phase of the process and associated with the probabilistic state transition rate $r_i$, $i = 1,2,...,p$. These transition rates govern the efficiency of DNA amplification. We derive the probability generating function \citep{pgf} for the number of target molecules found at an arbitrary time $t$. We use this generating function to derive the corresponding probability distribution and, importantly, the probability density function (pdf) of the $Ct$ value. We derive the pdf in two different cases, namely
        \begin{enumerate}
            \item when the initial state of the PCR process is deterministic, and the PCR phase lengths and amplification efficiencies are given; and
            \item when the initial state is Poisson-distributed, and the phase lengths and amplification efficiencies are given.
        \end{enumerate}
        To illustrate the mathematical ideas, we will report calculations and simulations based on a single-phase model. We argue that this simpler instance of our model is sufficient for analysing a large variety of real-world PCR experiments. In principle, each PCR experiment can be divided into the following three amplification rate-dependent phases: a pre-exponential phase, in which the amplification rate is sub-exponential; an exponential phase; and a post-exponential phase where the rate slows down as DNA molecules saturate the reagents required for their further amplification. However, in practice, the usual output of PCR experiments -- the $Ct$ value -- is determined as soon as the PCR process enters the exponential phase, meaning that dynamics occurring in the pre-exponential phase primarily determine this particular outcome. Therefore, for the purposes of understanding the factors that shape the $Ct$ value and its statistics, and evaluating related operating characteristics of PCR, a single-phase model appears sufficient. Accordingly, when applicable, we highlight the forms taken by our mathematical equations in the case of a single-phase model. In addition, we estimate the LoD and LoQ using a single-phase model (Section \ref{lodq}), which we also apply to critique the standard method of interpreting digital PCR data (Section \ref{dpcr}).
 \subsection{Case 1: A PCR process with a deterministic initial state}\label{n_known}
	\subsubsection{Probability generating function for the number of molecules}\label{p_gen1}
	\begin{theorem} \label{thm1}
		Let $\{X(t), t \in R\}$  be a continuous-time Markov process with $p$ phases, a countable state space $S \subset  \mathbb{N^+}$, phase-specific transition rates $r_i,$ $ i \in {1,2, \ldots ,p},$ and state transition probability given by 
		\begin{equation} \label{trans}
			P\left(X(t^\prime+\Delta t)=x|X(t^\prime)=x^\prime\right)= \delta (x^\prime-x+1) \sum_{i=1}^{p} r_i \mathbbm{1}_{I_i}(t^\prime),
		\end{equation}
		where $\mathbbm{1}$ denotes the indicator function and $\delta(.)$ denotes the Kronecker delta function.		
		If the process starts with $n$ molecules, then the probability generating function for the number of molecules present at time $t\in I_k$, $k \leq p$, is given by 
		\begin{equation}
			\label{thm11}
			G(n,\vec{r},t,\vec{\tau};s) = \left[\frac{se^{-z}}{1-s\left(1 - e^{-z}\right)}\right]^n,
		\end{equation} 
		where 
		\begin{equation}
		z=r_k t+\sum_{i=1}^{k-1}(r_i-r_{k})\tau_i,
		\label{zeq}
		\end{equation}
	$I_i$ denotes the i'th phase and $\tau_i = |I_i|$, $i < k$, is its length.
	\end{theorem}
	The proof of this theorem is given in Section \ref{pthm1}. We will now use the theorem to derive the probability distribution of the number of molecules found at time $t$.
	\subsubsection{Probability distribution of the number of molecules}\label{p_x1}
	\begin{corollary}\label{thm2}
		The probability that there are $x$ molecules at time $t\in I_k$ in the PCR process described in Theorem \ref{thm1} is given by the following negative binomial distribution:
		\begin{equation}
			P(x|n,\vec{r},t,\vec{\tau}) = \binom{x-1}{n-1}e^{-nz}\times\left(1-e^{-z}\right)^{x-n},
            \label{pdfx1}
		\end{equation}
		where $z$ is given by \eqref{zeq}.
	\end{corollary}
	The proof of this corollary is given in Section \ref{pthm2}. We will now use this corollary to derive the pdf, mean, variance and cdf of the $Ct$ value. 

	\subsubsection{pdf, mean and variance of the Ct value} \label{p_t1}
	Let $t$ be the $Ct$ value of the PCR process described in Theorem \ref{thm1}. By definition, $t$ is the time at which the number of molecules reaches the quantification threshold, which we denote by $x$. Let $t \in I_k$. In the Appendix [Section \ref{statsn}], we show that, given $n, \vec{r}=(r_1,r_2,\dots,r_{k-1})$, and $\vec{\tau}=(\tau_1,\tau_2,\dots,\tau_{k-1})$, the pdf of $t$ has the following form: 
			\begin{eqnarray}
		P(t|n,\vec{r},\vec{\tau},x) &=& \frac{r_k e^{-nz}\left(1-e^{-z}\right)^{x-n}}{B_{\theta}(n,x-n+1)},
		\label{pdf1t}
		\end{eqnarray} 
	where $B_{\theta}(n,x-n+1)$ is the incomplete Beta function, $z$ is given by \eqref{zeq}, and
  \begin{equation}
 \theta = e^{-\sum_{i=1}^{k-1}r_i\tau_i}. \label{theta_eqnt}
 \end{equation}
For the single-phase PCR process, $\theta = 1$, so the pdf is given by
		\begin{eqnarray}
		P(t|n,r_1,x) &=& \frac{r_1 e^{-nz}\left(1-e^{-z}\right)^{x-n}}{B(n,x-n+1)}.
		\label{pdf11t}
		\end{eqnarray}
	The mean $Ct$ value is given by (see Section \ref{statsn})
	\begin{eqnarray}
	\mathbb{E}(t) &=& \sum_{i=1}^{k-1}\tau_i + \frac{\Gamma(n)^2 {\theta}^n \: {}_3\tilde{F}_2(n,n,n-x;n+1,n+1;\theta)}{r_k B_{\theta}(n,x-n+1)},
   \label{mean1t}
	\end{eqnarray}
where ${}_3\tilde{F}_2(n,n,n-x;n+1,n+1;\theta)$ is the regularized generalized hypergeometric function.

For the single-phase process, the mean is given by
\begin{eqnarray}
\mathbb{E}(t) &=&  \frac{\psi(x+1)-\psi(n)}{r_1}.
\label{mean11t}
\end{eqnarray}
Observe that when $n \gg 1$, the right-hand-side of \eqref{mean11t} is well-approximated by $\ln(x/n)/r_1$. The latter expression is commonly used to approximate the mean $Ct$ value. For example, it was used in \citep{mutesa} to estimate PCR amplification efficiency from data.		

The variance of the $Ct$ value is given by $\mathbb{E}(t^2) - \mathbb{E}(t)^2$, where $\mathbb{E}(t^2)$ is given by \eqref{var1}.
For the single-phase process, the variance is given by
\begin{equation}
\textbf{Var}(t) = \frac{\psi_1(n) - \psi_1(x+1)}{r_1^2}, \label{var11t}
\end{equation}
where $\psi_1(\cdot)$ is the second polygamma function (also called the trigamma function).

Finally, the cdf of the $Ct$ value is given by [see Section \ref{statsn}]
		\begin{eqnarray}
			F(t|n,\vec{r},\vec{\tau},x) &=& 1 - \frac{B_{e^{-z}}(n,x-n+1)}{B_{\theta}(n,x-n+1)}.
			\label{cdf1t}
		\end{eqnarray}
For the single-phase process, the cdf is given by
\begin{eqnarray}
F(t|n,r_1,x) &=& 1 - I_{e^{-r_1t}}(n,x-n+1),
\label{cdf11t}
\end{eqnarray}
where $I_{e^{-r_1t}} (n, x-n+1) = B_{e^{-r_t}}(n,x-n+1)/B(n,x-n+1)$ is the regularized incomplete Beta function. Sampling from this cdf is relatively straightforward: A random $Ct$ value $t$ is obtained as follows:
\begin{eqnarray}
		t &=& -\frac{\ln I_{1-u}^{-1}(n, x-n+1)}{r_1},\label{samplingt}
	\end{eqnarray}
where  $u$ is sampled uniformly at random from the interval $(0,1)$ and  $I^{-1}_{1-u}$ is the inverse of the regularized incomplete Beta function. To find a $Ct$ value that corresponds to a quantile  $q \in (0,1)$, $q$ is substituted for  $u$.

To estimate $n$, either the pdf or the cdf of $t$ can be fit to data. Alternatively, the posterior density of $n$ conditioned on $t$ can be computed. In Section \ref{statsn}, we show that, for the single-phase process, it is given by
\begin{eqnarray}
P\left(n|r_1,t,x\right) &=& \frac{e^{-(n-1)r_1t}(1-e^{-r_1t})^{x-n}}{x B(n,x-n+1)}.
\label{pdfnt}
\end{eqnarray}

\begin{figure}
    \includegraphics[width=\textwidth]{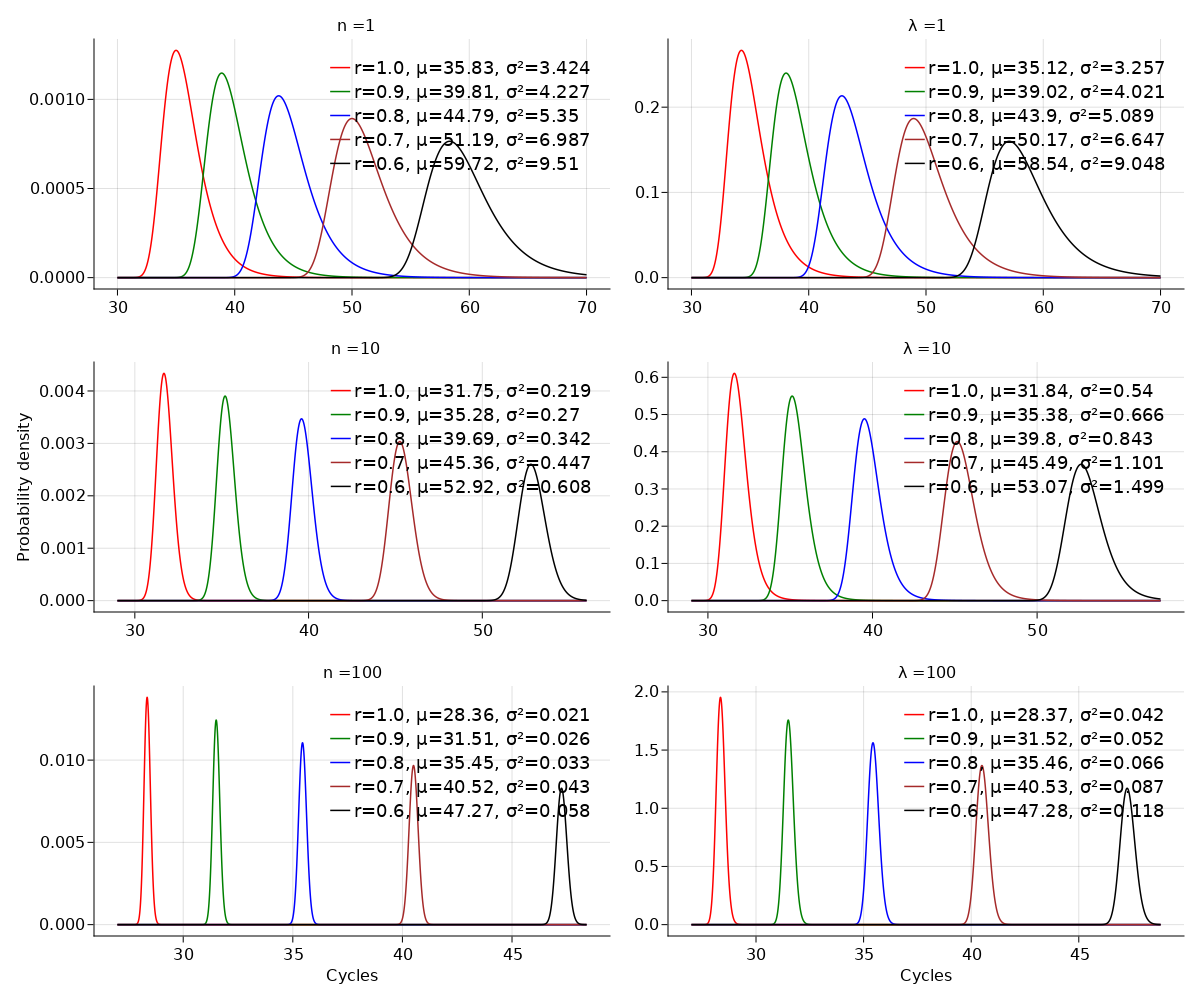}
    \centering
    \caption{\textbf{pdf of the quantification cycle for the single-phase process}. Processes with either a deterministic (\textit{left panel}) or a Poisson-distributed (\textit{right panel}) initial state were considered. The pdf was calculated using Equation \eqref{pdf1t} for the former case, and Equation \eqref{pdf2t} for the latter case. The quantification threshold, $x$, was set to $2^{35}$. The mean $\mu$ and variance $\sigma^2$ corresponding to different amplification efficiences $r$ are shown. For ease of comprehension, $r$, which in our model is defined on a base-$e$ scale, is shown as a percentage of its maximum possible value of $\ln(2)$.}
    \label{fig:fig1}
\end{figure}
	
	In Figure \ref{fig:fig1}, we illustrate the shape of the single-phase pdf for different numbers of input molecules and different amplification efficiencies. To this end, we set $T=35$ (a common upper-bound for the duration of real-world PCR experiments) and $x = 2^T$, which is equal to the number of molecules expected after $T$ cycles under perfect amplification conditions (a sample that contains only one, perfectly amplified input molecule is expected to reach the quantification threshold, $x$, at time $t \leq T$). We vary the efficiency from 60\% m.p.e. (equivalent to setting $r_1 = 0.6\times\ln2$) to 100\% m.p.e. The pdf has a bell shape, the location and width of which are governed by both the efficiency and the number of input molecules (Figure \ref{fig:fig1}, left panel). Higher efficiencies or larger numbers of input molecules produce smaller mean $Ct$ values, smaller variances, and narrower pdfs. In contrast, lower efficiencies or smaller numbers of input molecules produce larger mean $Ct$ values, larger variances, and wider pdfs (Figure \ref{fig:fig1}, left panel). In fact, Equation  \eqref{pdf1t} predicts that in the limit as the efficiency goes to 0, the pdf will become flat as it will map every $Ct$ value to 0.

	\subsection{Case 2: A PCR process with a Poisson-distributed initial state}\label{lam_known}
	\subsubsection{Probability generating function for the number of molecules}\label{p_gen2}
	\begin{theorem} \label{thm3}
		Let $\{X(t), t \in R\}$  be the continuous-time Markov process described in Theorem \ref{thm1}. If, instead of starting with a precisely known number of input DNA molecules, the initial state of the process is Poisson-distributed with mean $\lambda$, then the probability generating function for the state of the process at a future time $t\in I_k$ is given by 
		\begin{equation}
			\label{thm21}
			G(\lambda,\vec{r},t,\vec{\tau};s) = e^{\frac{\lambda(s-1)}{1-s\left(1 - e^{-z}\right)}},
		\end{equation} 
		where $z$ is given by \eqref{zeq}.
	\end{theorem}
	The proof of  Theorem \ref{thm3} is given in Section \ref{pthm3}.
	We now use Theorem \ref{thm3} to derive the probability distribution of the number of molecules found in the PCR process at time $t$.
	\subsubsection{Probability distribution of the number of molecules}\label{p_x2}
	\begin{corollary}\label{thm4}
		The probability that there are $x$ molecules at cycle $t\in I_k$ in the PCR process described in Theorem \ref{thm3} is given by:
		\begin{equation}
			P(x|\lambda,\vec{r},t,\vec{\tau}) =  e^{-\lambda} \left(1-e^{-z}\right)^{x}\sum_{i=1}^{x}\frac{\binom{x-1}{i-1}}{i!} \left(\frac{ \lambda e^{-z} }{1 - e^{-z}} \right)^{i},
            \label{pdfx2}
		\end{equation}
		where $z$ is given by \eqref{zeq}.
	\end{corollary}
	The proof of this corollary is provided in Section \ref{pthm4}. It is interesting to note that from the proof emerged the following combinatorial triangle, which is related to the well-known Narayana triangle \citep{narayana}:\newline
		\begin{center}
			\begin{tabular}{>{}l<{}|*{6}{c}}
				\multicolumn{1}{l}{$x$} &&&&&&\\\cline{1-1} 
				1 &1&&&&&\\
				2 &1&2&&&&\\
				3 &1&6&6&&&\\
				4 &1&12&36&24&&\\
				5 &1&20&120&240&120&\\\hline
				\multicolumn{1}{l}{} &1&2&3&4&5\\\cline{2-7}
				\multicolumn{1}{l}{} &\multicolumn{6}{c}{$k$}.
			\end{tabular}
		\end{center}
		The entries of this triangle, given by 
		\begin{equation}
			T(x,k)= \binom{x}{k-1}\binom{x-1}{k-1}\:(k-1)!, \:\: x \in \mathbb{Z}^{+}, k=1,2,...,x,
		\end{equation}
count the number of ways of obtaining $x$ molecules by replicating a randomly selected subset of $k$ molecules. $T(x,k)$ is related to the Narayana numbers $N(x,k)$ by
$$T(x,k) = k! \: N(x,k).$$
	
	We will now use this corollary to derive the pdf of the $Ct$ value together with the mean, variance and cdf.
	
	\subsubsection{pdf, mean and variance of the $Ct$ value}\label{p_t2}
	Let $t$ be the $Ct$ value of the PCR process described in Theorem \ref{thm3}. As noted earlier, the $Ct$ value $t$ is the time at which the number of DNA molecules found in the process reaches the quantification threshold, which we denote by $x$. In the Appendix [Section \ref{statslam}], we show that the pdf of $t$ is given by
	\begin{eqnarray}
			P(t|\lambda, \vec{r}, \vec{\tau},x) &=& \frac{r_k \lambda e^{-z} (1 - e^{-z})^{x-1} \: {}_1F_1 \left(1-x;2; \frac{-\lambda e^{-z}}{1 - e^{-z}} \right)} {\sum_{j=1}^{x} \frac{\binom{x-1}{j-1} \lambda ^j}{j!} B_{\theta}(j,x-j+1)},
		\label{pdf2t}
			\end{eqnarray}
where $\theta$ is given by \eqref{theta_eqnt}, $z$ is given by \eqref{zeq}, ${}_1F_1(a;b;c)$ is the hypergeometric function (also called the Kummer confluent hypergeometric function of the first kind).

For the single-phase process, the pdf is given by [see Section \ref{statslam}]
\begin{equation}
	P(t|\lambda, r_1,x) = \frac{r_1 x \lambda e^{-r_1t} (1 - e^{-r_1t})^{x-1} \: {}_1F_1 \left(1-x;2; \frac{-\lambda e^{-r_1t}}{1 - e^{-r_1t}} \right)} {e^{\lambda} - 1}.
			\label{pdf21t}
\end{equation}

 The mean $Ct$ value is given by [see Section \ref{statslam}]
		\begin{eqnarray}
			\mathbb{E}(t)&=& \frac{\sum_{j=1}^{x} \frac{\binom{x-1}{j-1} \lambda ^j}{j!} \Big[r_kB_{\theta}(j,x-j+1) \sum_{i=1}^{k-1}\tau_i + \Gamma(j)^2 {\theta}^j \: {}_3\tilde{F}_2(j,j,j-x;j+1,j+1;\theta) \Big]}{r_k \sum_{j=1}^{x} \frac{\binom{x-1}{j-1} \lambda ^j}{j!} B_{\theta}(j,x-j+1)},
			\label{mean2t} 
		\end{eqnarray}
while the second moment is given by \eqref{var2}.

For the single-phase process, the mean and variance are, respectively, given by [see Section \ref{statslam}]
\begin{eqnarray}
	\mathbb{E}(t) &=& \frac{ \psi(x+1)}{r_1} - \frac{\sum_{j=1}^{x} \frac{\lambda ^j }{j!} \psi(j)}{r_1 \left(e^{\lambda}-1 \right)} \text{\:\: and}
	\label{mean21t}
\end{eqnarray}
\begin{equation}
    \textbf{Var}(t) =  \frac{ \left(e^{\lambda}-1 \right) \sum_{j=1}^{x} \frac{\lambda^j}{j!}  \Big[\psi_1(j) + \psi(j)^2 \Big] - \Big[ \sum_{j=1}^{x} \frac{\lambda^j}{j!}  \psi(j) \Big]^2}{\left(r_1 (e^{\lambda}-1) \right)^2} - \frac{\psi_1(x+1)}{r_1^2}. \label{var21t}
\end{equation}
	The cdf of the $Ct$ value is given by [see Section \ref{statslam}]
	\begin{eqnarray}
	F(t|\lambda,\vec{r},\vec{\tau},x) &=& 1 - \frac{\sum_{j=1}^{x} \frac{\binom{x-1}{j-1} \lambda ^j}{j!} B_{e^{-z}}(j,x-j+1)}{\sum_{j=1}^{x} \frac{\binom{x-1}{j-1} \lambda ^j}{j!} B_{\theta}(j,x-j+1)}. \label{cdf2t}
\end{eqnarray}
For the single-phase process, the cdf is given by
\begin{eqnarray}
	F(t|\lambda,r_1,x) &=& 1 - \frac{x \sum_{j=1}^{x} \frac{\binom{x-1}{j-1} \lambda ^j}{j!} B_{e^{-r_1t}}(j,x-j+1)}{e^{\lambda}-1}.
		\label{cdf21t}
\end{eqnarray}
To estimate $\lambda$, either the pdf or the cdf of $t$ can be fit to data. Alternatively, the posterior density of $\lambda$ conditioned on $t$ can be computed. In Section \ref{statslam}, we show that, for the single-phase process, it is given by
\begin{equation}
P(\lambda|r_1,t,x) = \frac{\lambda w \; {}_1F_1(1-x;2;\frac{-\lambda w}{1-w})}{(e^{\lambda}-1)(1-w) \sum_{j=1}^{x} \binom{x-1}{j-1} \left(\frac{w}{1-w} \right)^j \zeta(j+1) },
	\label{pdflamt}
\end{equation}
where $w = e^{-r_1t}$ and $\zeta(\cdot)$ denotes the Riemann zeta function.

Note that, because they result from calculating expectations over the Poisson distribution, the summations found in Equations \eqref{pdf2t} - \eqref{pdflamt} can be truncated at any value of $j \gg \lambda$ without a loss of accuracy.

In  Figure \ref{fig:fig1}, we illustrate the shape of the single-phase pdf for different values of $\lambda$ and different amplification efficiencies. As was the case for the PCR process with a deterministic initial state (Figure \ref{fig:fig1}, left panel), the pdf also has a bell shape (Figure \ref{fig:fig1}, right panel). Its location and width are governed by both the efficiency and $\lambda$. Consistent with expectations, higher efficiencies or larger values of $\lambda$ produce smaller mean $Ct$ values, smaller variances, and narrower pdfs (Figure \ref{fig:fig1}, right panel). In contrast, lower efficiencies or smaller values of $\lambda$ produce larger mean $Ct$ values, larger variances, and wider pdfs.

\subsection{Limit of detection and limit of quantification}\label{lodq}
    We will now demonstrate theoretically how the mathematical framework we have developed can be applied to achieve certain operationally important objectives. In particular, it is often of interest to quantify the limit of detection (LoD) of a particular instance of the PCR method (henceforth referred to as ``PCR protocol''). The LoD of a PCR protocol is the smallest number of molecules that it can detect with a failure rate not exceeding a defined threshold $\alpha$ ($\alpha$ is also called the significance level). Protocols with smaller LoDs are in general preferred to those with larger LoDs. Ideally, the LoD should be either equal to or smaller than the number of input DNA molecules expected in the considered sample. Another important operational objective is to determine a PCR protocol's limit of quantification (LoQ) -- i.e. the smallest number of molecules that it can estimate with a given level of precision (measured here using the parameter $\beta$) and a given maximum failure rate $\alpha$. When a $\geq \beta$-fold change in the number of target molecules needs to be detected, the protocol should have an LoQ with precision $\leq \beta$. The methods available for estimating LoD and LoQ are laborious \citep{forootan2017methods} and frequently rely on certain \textit{ad-hoc} mathematical approximations \citep{forootan2017methods,nutz2011determination}, which we would like to circumvent by developing and executing mathematically precise statements of the estimation problem.
    
    We begin with the LoD estimation problem. For the PCR process with a deterministic initial state, the LoD can be expressed as follows:
    \begin{eqnarray}
		\text{LoD} = &\textbf{min} \:\: n \nonumber\\ &\text{s.t.} \:\: F(T|n, \vec{r}, \vec{\tau},x) > 1-\alpha,	\label{lodt}
	\end{eqnarray}
where $F(T|n, \vec{r}, \vec{\tau},x)$ is given by \eqref{cdf1t} and $T$ is the maximum practical duration of the PCR process. For the process with a Poisson-distributed initial state, $n$ is replaced by $\lambda$.

In Supplementary Figure \ref{fig:appdxfig2}, we show how the LoD varies with amplification efficiency in a single-phase process with either a deterministic or a Poisson-distributed initial state. In the former case, the process contains amplification noise but no sampling noise while in the latter case it contains both sampling noise and amplification noise. For comparison, we also show the LoD in a process with sampling noise, modeled by using the Poisson distribution, but without amplification noise. The LoD is lowest in a process with sampling noise alone (LoD = 3 molecules) and it is highest when both sampling noise and amplification noise are present (LoD ranges from 6 molecules, at 100\% of maximum possible efficiency or m.p.e, to 157 molecules, at 80\% m.p.e). A process with amplification noise but without sampling noise has an intermediate LoD. In these computational examples, the parameters of the equation used to estimate the LoQ are perfectly known, and this makes it possible to obtain perfect knowledge of the LoD. In real-world applications, the parameter values will be associated with uncertainty, which will, in a quantifiable way, make uncertain the LoD estimates. 

We now turn our attention to the problem of estimating the LoQ in a PCR process with a deterministic initial state. Suppose that a $Ct$ value $t$ is generated by such a process and then used to obtain an estimate, denoted $\hat{n}$, of the number of input molecules. Let $n$ be the actual number of input molecules. We want to calculate the probability that, for any data $t$ generated by the same process, $\hat{n}$ will not differ from ${n}$ by more than a factor $\beta, \beta \geq 1$. We define the LoQ as the smallest value of $n$ for which this probability exceeds $1-\alpha$. Specifically, the LoQ is given by
		\begin{eqnarray}
		\text{LoQ} = &\textbf{min} \:\: n \: \nonumber\\ &\text{s.t.} \:\: P\left(\left \lfloor n/\beta \right \rfloor \leq \hat{n} \leq \left \lceil \beta n \right \rceil \:| \: n,\vec{r},\vec{\tau},x\right) > 1-\alpha,
	\label{loqt}
	\end{eqnarray}	
 where $\left \lfloor v\right \rfloor$ (respectively $\left \lceil v \right \rceil$) denotes the largest (respectively smallest) integer less than (respectively greater than) or equal to $v$.

 Focusing on the single-phase process, to obtain $P\left(\left \lfloor n/\beta \right \rfloor \leq \hat{n} \leq \left \lceil \beta n \right \rceil \:| \: n,r_1,x\right)$, we marginalize the right-hand-side of \eqref{pdfnt} with respect to $t$ and then take the sum from $\left \lfloor n/\beta \right \rfloor$ to $\left \lceil \beta n \right\rceil$, yielding [see Section \ref{statsn}]
\begin{eqnarray}
P\left(\left \lfloor n/\beta \right \rfloor \leq \hat{n} \leq \left \lceil \beta n \right \rceil \:| \: n,r_1,x\right) &=& \sum_{\hat{n}=\left \lfloor n/\beta \right \rfloor}^{\left \lceil \beta n \right \rceil} \frac{B(\hat{n}+n-1,2x-\hat{n}-n+1)}{x B(\hat{n},x-\hat{n}+1)B(n,x-n+1)} \nonumber\\
&=& P\left(\left \lfloor n/\beta \right \rfloor \leq \hat{n} \leq \left \lceil \beta n \right \rceil \:| \: n,x\right).
\label{liket2}
\end{eqnarray}
Strikingly, while Equation \eqref{liket2} depends on both $n$ and $x$, it does not depend on the amplification efficiency $r_1$. In the Appendix [see Equation \eqref{lamhat}], we follow a similar procedure to obtain $P\left(\left \lfloor \lambda/\beta \right \rfloor \leq \hat{\lambda} \leq \left \lceil \beta \lambda \right \rceil \:| \: \lambda,r_1,x\right)$, the probability that, for any data $t$ generated by a single-phase PCR process with a Poisson-distributed initial state, the estimated value of${\lambda}$, denoted $\hat{\lambda}$, will not differ from the actual value by more than a factor $\beta$. 
	
Setting $\alpha = 0.05$ and allowing at most a 10\% deviation of $\hat{n}$ from $n$ (corresponding to setting $\beta = 1.1$) results in an LoQ of 820 molecules. The LoQ decreases to 146 molecules when a deviation of up to 25\% from expectation is allowed (corresponding to $\beta = 1.25$), and to 43 molecules when the allowable deviation increases to 50\% (corresponding to $\beta = 1.5$). The analysis suggests that at the considered 5\% failure rate, 10 input molecules can be detected with an error of at least $\approx$200\%, that is, a $\approx 2$ fold deviation from $n$ (corresponding to $\beta = 2$). Because these calculations do not account for sampling noise, they provide only a lower-bound for the LoQ that is achievable at the considered failure rate and level of precision ($\beta$). As noted earlier, in real-world applications the parameters of the equation used to estimate LoQ will be imperfectly known, and this will determine the amount of uncertainty associated with the estimated LoQ. 
	\subsection{Amplification noise determines digital PCR outcomes}\label{dpcr}
As noted earlier, digital PCR is a variant of conventional PCR that was developed to improve the quantification of DNA. In digital PCR, a sample master mix (containing an unknown number of input DNA molecules together with all the reagents required for DNA replication) is uniformly distributed into hundreds (and sometimes thousands) of physical partitions, which may take the form of droplets or microwells \citep{dpcr0}. Each partition is expected to receive zero, one, or more DNA molecules following a Poisson distribution with mean $\lambda = C V / D$, where $C$ is the concentration of the DNA in the original sample, $V$ is the partition volume, and $D$ is the (known) dilution factor applied to the sample during preparation of the mastermix. PCR reactions are independently and simultaneously run inside each partition, and positive partitions are identified. The resulting data -- i.e. the positive or negative outcome of each PCR reaction -- are thus digital. The standard method of interpreting these data assumes that partitions that receive at least one target molecule will test positive, and their fraction, $\hat{f}$, is approximated by $\hat{f} \approx 1- e^{-{\lambda}}$, from which ${\lambda}$ is estimated as $\hat{\lambda} = -\ln(1-\hat{f})$ and then used to estimate $C$.
	
	However, according to our model, the assumption that the fraction of positive partitions equals the Poisson probability that a partition receives one or more target molecules is untenable due to the effects of PCR amplification noise. Indeed, setting the amplification efficiency to a reasonably high value of 95\% m.p.e (i.e. $r_1=0.95 \; \ln2)$ and using $T=35, x=2^T$ in Equation \eqref{cdf21t}, we predict that $<$4\%  of partitions that contain only one molecule will test positive, which is $>$25-fold smaller than assumed by the Poisson method. In Supplementary Figure \ref{fig:appdxfig3}, we compare the fraction of positive partitions calculated using our model [Equation\eqref{cdf21t}] versus the positive fraction calculated by the Poisson method, for different values of $\lambda$ and different amplification efficiencies. We find that the Poisson method over-estimates the fraction of positive partitions for small values of $\lambda$, including the value (1.61) at which the method is expected \citep{dpcr3} to produce its most precise estimates of $\lambda$. Only for a relatively large value of $\lambda$ (10) do we find the Poisson method's estimate of the fraction of positive partitions to agree with the noise-adjusted expectation calculated using our model (Supplementary Figure \ref{fig:appdxfig3}).
\begin{figure}
	\centering
		\includegraphics[width=\textwidth]{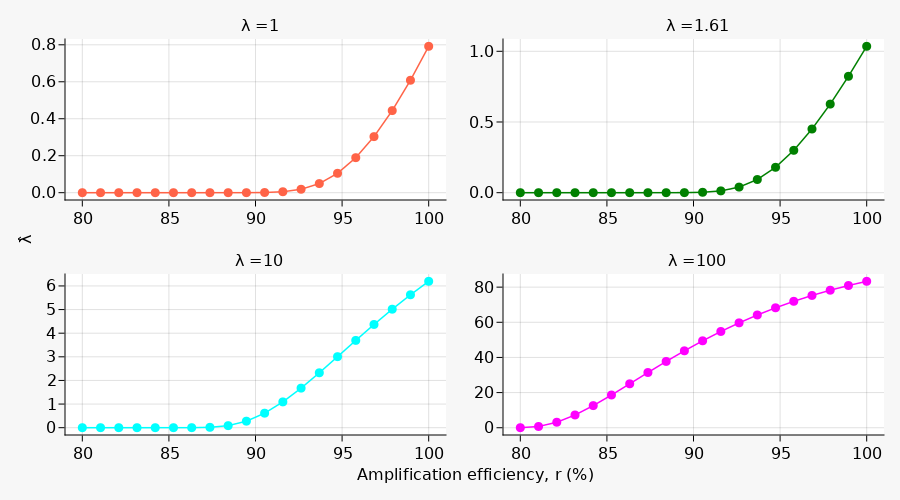}
\caption{ \textbf{Under-estimation of $\lambda$ by the standard method of interpreting digital PCR data.} We varied both the amplification efficiency and the amount of input DNA $\lambda$ and used Equation \eqref{lamt} to estimate $\lambda$. The resulting estimate, denoted $\hat{\lambda}$, was plotted against efficiency, expressed as a percentage of the maximum possible efficiency. At efficiencies lower than 95\%, only a very small amount of the input DNA is detected. At 95\% efficiency, the amount detected ranges from 12\%, when $\lambda=1$, to 69\%, when $\lambda=100$. The amount detected increases to $\approx$80\% when the efficiency equals 100\%. }
\label{fig:fig4}
\end{figure}

	We use our model to investigate how this over-estimation of the fraction of positive partitions affects the accuracy of the estimate of $\lambda$ (denoted $\hat{\lambda}$) produced by the Poisson method. Setting $t=T$ in Equation \eqref{cdf21t}, we find that
	\begin{eqnarray}
	\hat{\lambda} &=& -\ln\left(1-\hat{f}\right) \nonumber\\
        &=& \ln\Big( \frac{e^{\lambda} - 1}{ \sum_{j=1}^{x} \frac{\binom{x}{j}}{(j-1)!} \lambda ^j B_{e^{-r_1T}}(j,x-j+1)} \Big).
	\label{lamt}
\end{eqnarray}
	According to Equation \eqref{lamt}, $\hat{\lambda}$ depends strongly on the amplification efficiency $r_1$. $\hat{\lambda}$ equals 0 in the limit as $r_1$ tends to 0. As $r_1$ increases to its maximum possible value of $\ln(2)$, $\hat{\lambda}$ also increases, approaching $\lambda$. In Figure \ref{fig:fig4}, we illustrate the relationship between $\hat{\lambda} /\lambda$ and amplification efficiency. Strikingly, for efficiencies lower than 95\% m.p.e, $\hat{\lambda} / \lambda$ is very small, indicating that $\lambda$ is markedly under-estimated by the Poisson method. When the efficiency equals 95\% m.p.e, $\hat{\lambda}/\lambda$ increases from $\approx 12\%$  (at $\lambda=1$) to $\approx$69\% (at $\lambda=100$). Increasing the efficiency to the maximum possible value of 100\% m.p.e causes $\hat{\lambda} /\lambda$ to increase to $\approx$80\% (at $\lambda=100$). These results indicate that the Poisson method is expected to under-estimate $\lambda$ because it does not account for amplification noise. Indeed, experimental data show a strong tendency by the method to under-estimate the number of input DNA molecules (eg. see Supplementary Table 6 in \citep{hindson}).
	
	\section{Discussion}\label{discuss}
	The outputs of DNA quantification experiments, including those based on the polymerase chain reaction (PCR), tend to vary within and across different experimental instances, making the results difficult to interpret and limiting their utility beyond the particular contexts in which they are generated. Indeed, various factors are known to contribute to the variability of PCR outputs \citep{qpcr2,additives,review1} including the varying complexity of DNA templates and the random distribution of target molecules in the reaction environment; the type of PCR machine and buffer components used; the durations and temperatures of the three thermal cycles of PCR; the binding kinetics of oligonucleotide primers to target DNA; and the stability of DNA polymerase  and other PCR reagents. Taylor et. al \citep{review1} reviewed the sources of variability in PCR experiments and proposed a stepwise process to minimize such variability in practice.
	
	A common output of a PCR experiment is the quantification cycle (denoted $Ct$ or $Cq$ value), the PCR cycle at which the number of DNA molecules exceeds a defined threshold, called the quantification threshold. The $Ct$ value varies with both the number of input DNA molecules and the PCR amplification efficiency, which in turn varies with the aforementioned experimental variables. It is desirable to deconvolute such variable outputs to estimate the number of input DNA molecules, which is of greatest interest in experiments, by applying mathematical methods that account for the stochasticity that is inherent in the underlying generative process.	
	
	We have developed a mathematical approach to modeling DNA quantification that takes into account the underlying stochasticity. We used PCR, the most widely used class of DNA quantification process, to illustrate our mathematical ideas, which are also applicable to a broader class of such processes (eg. \citep{lamp}). Using the model, we derived the probability generating function for the number of molecules found in a PCR process with either a deterministic or a Poisson-distributed number of input molecules as well as the probability density function (pdf), mean, variance and cumulative density function (cdf) of the $Ct$ value produced by such a process. In contrast to the deterministic case, in which PCR outputs are contaminated only by amplification noise, in the latter case the outputs also contain sampling noise. The equations we derived for these important statistical properties of the PCR process revealed functional relationships between the $Ct$ value and underlying variables that could previously only be accessed by empirical means.
	
	To illustrate our mathematical ideas, we focused on the single-phase PCR process, for which our modeling results take relatively simple mathematical forms. We found that the common assumption that the mean $Ct$ value is a simple logarithmic function of the number of input DNA molecules $n$ is correct when $n$ is large. For small $n$, corrections are required. An exact mean $Ct$ value is given by $\left(\psi(x+1)-\psi(n)\right)/r$, where $x$ is the quantification threshold, $r$ is the amplification efficiency and $\psi(\cdot)$ denotes the first polygamma function. The variance has the elegant form $\left(\psi_1(n)-\psi_1(x+1)\right)/r^2$, where $\psi_1(\cdot)$ denotes the second polygamma function. Therefore, the variance is strongly dependent on amplification efficiency. This effect is illustrated in Figure \ref{fig:fig1}, which shows that the pdf of the $Ct$ value becomes wider as the amplification efficiency decreases. Interestingly, in this simple case, the coefficient of variation of the $Ct$ value (i.e. the ratio of the standard deviation to the mean) does not depend on amplification efficiency.
	
Two important numbers that characterize the performance of a PCR process are the limit of detection (LoD) and the limit of quantification (LoQ). The LoD is the smallest number of molecules that can be detected with a failure rate not exceeding a threshold $\alpha$, while LoQ is the smallest number of molecules that can be quantified with a given level of precision (i.e. allowing a defined maximum fold deviation from the true value) and a given maximum failure rate $\alpha$. We provided mathematical formulae for calculating both LoD and LoQ. Close examination of these formulae in the context of a single-phase PCR process revealed that a small reduction of the amplification efficiency may cause a large increase of LoD. For example, when $\alpha = 5\%$, reducing the efficiency from 95\% of the maximum possible efficiency (m.p.e) to 90\% m.p.e. caused the LoD to double, from $\approx$10 input molecules to $\approx$20 molecules. In contrast to LoD, we found that LoQ is independent of efficiency. Allowing up to a 2-fold difference between $n$ and its estimate results in an LoQ of $\approx$10 molecules. Reducing the allowed fold difference to 10\% increases LoQ to 820 molecules. Our methods may be used to improve significantly the current approaches to estimating LoD and LoQ, which are laborious \citep{forootan2017methods} and frequently rely on certain crude mathematical approximations \citep{forootan2017methods,nutz2011determination} that can be avoided by using our methods.

Furthermore, we applied our methods to shed light on the effects that amplification noise has on estimates of the expected number of input DNA molecules $\lambda$ obtained by the standard method of interpreting digital PCR data. A key assumption of this method is that a PCR reaction will be positive if it contains at least one input DNA molecule. We showed that this assumption is in general invalid because of the stochastic nature of PCR amplification. Stochastic effects are particularly large when $\lambda$ is small, which is the regime in which digital PCR preferentially operates. At a high amplification efficiency of 95\% m.p.e, we find that the ratio of the fraction of positive digital PCR reactions calculated by the standard method versus the value obtained after accounting for amplification noise is only 18.3\% when $\lambda=1$ and it increases to $\approx$100\% when $\lambda=10$ (Figure \ref{fig:appdxfig3}). Accordingly, stochastic effects were found to cause a significant under-estimation of $\lambda$ by the standard method. Indeed, at a high efficiency of 95\% m.p.e., the standard method is predicted to under-estimate $\lambda$ by factors of $\approx$8.1, $\approx$3.1, and $\approx$1.4 when $\lambda$ equals 1, 10, and 100, respectively (Figure \ref{fig:fig4}). This is in the same range as empirically observed (eg. \citep{hindson}).

    Using our mathematical methods, the following two different approaches may be used to obtain much more accurate estimates of $\lambda$. Firstly, if $Ct$ values are available from positive digital PCR reactions, then Equation \eqref{pdf21t} can be fit to those $Ct$ values, using either a likelihood-based or a Bayesian statistical approach, to estimate the most probable value of $\lambda$ together with a confidence (or credible) interval for it. Secondly, if only binary (ie. positive or negative) outcomes are available from individual reactions, then the variability of such outcomes can still be exploited to estimate $\lambda$. Specifically, suppose there are $N$ different reactions. These can be randomly distributed into groups of $N'$ reactions each. Assuming a binomial distribution of the number of positive reactions found in each group, their first and second moments are given by $F(T)N'$ and $F(T)N'\left(1+F(T)(N'-1)\right)$,  respectively, where $F(T)$ is calculated using \eqref{cdf21t}. These moments contain information about the two free parameters of $F(T)$ (i.e. $\lambda$ and $r_1$), which can be readily extracted to estimate $\lambda$ together with a confidence (or credible) interval.   
		
	The $Ct$ value is estimated from the fluorescence profiles produced by DNA molecules as they are amplified during PCR. Our mathematical analysis can be straightforwardly extended to obtain a time-dependent probability density of the fluorescence intensity $P_t(y)$, which can then be fit to fluorescence profiles as an alternative approach to estimating the number of input DNA molecules. Using standard results from probability theory (eg. see \citep{siegrist}), $P_t(y)$ can be derived from both the cumulative distribution function of the number of molecules found in the PCR process at time $t$, $F_t(x)$ [calculated based on Equation \eqref{pdfx1} or \eqref{pdfx2}] and the linear relation expected \citep{rengarajan} between $y$ and $x$. Specifically, $P_t(y)$ can be expressed as   	
	 
	\begin{equation}
		P_t(y) = \frac{1}{\alpha} h_t(g^{-1} (y))   \label{dcdf},
	\end{equation}
where	
	\begin{equation}
		h_t(x) = \frac{d}{dx} F_t(x)   \label{dcdf2},
	\end{equation}
	$y = \alpha x + \beta = g(x)$, and $\alpha, \beta > 0 $. We will explore in detail this alternative approach to estimating the number of input DNA molecules in a future paper.

\renewcommand\thefigure{\thesection.\arabic{figure}}    
\section{Supporting Information}
\setcounter{figure}{0}    

	\subsection{Appendix}\label{appendix}
	This section contains mathematical proofs and detailed calculations supporting the results presented in Section \ref{res}. 
	\subsubsection{Proof of Theorem \ref{thm1}} \label{pthm1}
	\begin{proof}  We will prove Theorem \ref{thm1} by mathematical induction on $k$.
		\begin{itemize}
			\item $k=1$: \\ The Chapman-Kolmogorov forward equation corresponding to the single-phase process is given by:  
			\begin{equation}\label{me}
				\frac{\partial P(X=x,t|X=x^\prime, t^\prime)}{\partial t} = r_1(x-1)P(X=x-1,t|X=x^\prime, t^\prime)-r_1xP(X=x,t|X=x^\prime, t^\prime),
			\end{equation}
			where we have set $t = t^\prime + \Delta t$, and $r_1$ is the amplification efficiency associated with the process. To simplify our notation, we will abbreviate $P(X=x,t|X=x^\prime, t^\prime)$ by $P(x,t)$.
			
			We will solve \eqref{me} by using a powerful combinatorial device called the probability generating function (pgf) \citep{pgf}. Recall that the pgf of $P(x,t)$ is defined as:
			\begin{equation}
				\label{pgf}
				G(s,t) = \sum_{x=0}^{\infty}s^xP(x,t),\nonumber
			\end{equation}
            where $s$ is a book-keeping variable.
            
			Multiplying both sides of  \eqref{me} by $s^x$ and summing over all possible values of $x$ yields:
			\begin{eqnarray}
				\sum_{x=0}^{\infty} s^x\frac{\partial P(x,t)}{\partial t} &=& r_1\sum_{x=0}^{\infty}(x-1)s^xP(x-1,t) -r_1\sum_{x=0}^{\infty}xs^xP(x,t)\nonumber\\
				&=& r_1s^2\sum_{x=0}^{\infty}(x-1)s^{x-2}P(x-1,t) - r_1s\sum_{x=0}^{\infty}xs^{x-1}P(x,t)\nonumber\\
				&=& = r_1s\left[s\sum_{x=0}^{\infty}(x-1)s^{x-2}P(x-1,t) - \sum_{x=0}^{\infty}xs^{x-1}P(x,t)\right].\label{part3}
			\end{eqnarray}
			Using
			\begin{eqnarray}
				\frac{\partial G(s,t)}{\partial s} &=& \sum_{x=0}^{\infty}xs^{x-1}P(x,t) \; \textrm{and} \nonumber \\
				\frac{\partial G(s,t)}{\partial t} &=& \sum_{x=0}^{\infty}s^{x}\frac{\partial P(x,t)}{\partial t}, \label{part2}
			\end{eqnarray}
			we simplify  \eqref{part3} to obtain
			\begin{equation}
				\label{pde}
				\frac{\partial G(s,t)}{\partial t} = r_1s(s-1)\frac{\partial G(s,t)}{\partial s},
			\end{equation}
			which is a partial differential equation (pde) in $G(s,t)$. 
            
            We will solve \eqref{pde} by the method of characteristics. To this end, we define new variables $$u = u(s,t) \: \text{and}\: v = v(s,t),$$ which will transform \eqref{pde} into the simpler equation
            \begin{equation}
                \frac{\partial W(u,v)}{\partial u} + H(u,v)W(u,v) = F(u,v),
                \label{gensol}
            \end{equation}
            which has the solution $$W(u,v) = e^{-\int H(u,v) du} \left[\int F(u,v) e^{\int H(u,v)du} + \varPsi(v) \right],$$
            where $$W(u,v) = G\left(s(u,v),t(u,v)\right).$$

            This requires that $v(s,t) = c$, where $c$ is an arbitrary constant. The resulting characteristic equation is given by 
            $$\frac{ds}{dt} = -r_1s(s-1),$$ 
            which has the solution
            $$\frac{s-1}{s}e^{r_1t} = c = v(s,t).$$
            
            Setting $u(s,t) = t$, we obtain
            \begin{eqnarray}
                \label{partials1}
                \frac{\partial G}{\partial t} &=& \frac{\partial W}{\partial t} = \frac{\partial W}{\partial u} \frac{\partial u}{\partial t} + \frac{\partial W}{\partial v} \frac{\partial v}{\partial t} \nonumber\\
                &=& \frac{\partial W}{\partial u} + \frac{r_1(s-1)}{s}e^{r_1t} \frac{\partial W}{\partial v}
            \end{eqnarray}
            and
            \begin{eqnarray}
                \label{partials2}
                \frac{\partial G}{\partial s} &=& \frac{\partial W}{\partial u} \frac{\partial u}{\partial s} + \frac{\partial W}{\partial v} \frac{\partial v}{\partial s} \nonumber\\
                &=& \frac{1}{s^2}e^{r_1t} \frac{\partial W}{\partial v}.
            \end{eqnarray}
            Substituting \eqref{partials1} and \eqref{partials2} into \eqref{pde} gives
            \begin{equation}
                 \frac{\partial W}{\partial u} = 0,
                 \label{gensol2}
            \end{equation}
            which has the same form as \eqref{gensol}. The solution to \eqref{gensol2} is given by
            \begin{eqnarray}
				\label{erf}
                & W(u,v) &= \varPsi(v) \nonumber\\
                \implies& G(s,t) &= \varPsi\left(\frac{s-1}{s}e^{r_1t}\right). 
            \end{eqnarray}
			If there are $n$ molecules at the start of the process ($t=0$), then $p(x,0)= 1$ if $x=n$ and $p(x,0)= 0$ otherwise. Therefore, 
			\begin{equation}
				\label{g0}
				G(s,0) = \varPsi\left(\frac{s-1}{s}\right) = \sum_{x=0}^{\infty} s^xP(x,0)= s^n.
			\end{equation}
			In \eqref{g0}, the argument $y$ of $\varPsi\left(y\right)$ maps onto $(\frac{1}{1-y})^n$, implying that
			\begin{equation}
				G(s,t) = \varPsi\left(\frac{s-1}{s}e^{r_1t}\right)
				=\left( \frac{1} {1 - \frac{s-1}{s}e^{r_1t}} \right)^n
				=\frac{s^ne^{-nr_1t}}{\left[1-s\left(1-e^{-r_1t}\right)\right]^n}.				\label{pde12}
			\end{equation}
			
			Equation \eqref{pde12} matches  \eqref{thm11} when $k=1$.
			
			\begin{corollary} \label{cor1}
				Equation \eqref{pde12} solves \eqref{pde}. 
			\end{corollary}
			\begin{proof}
				The right-hand-side of \eqref{pde} is
				\begin{eqnarray}
					\frac{\partial G}{\partial s}&=& ns^{n-1}e^{-nr_1t}\left[1-s(1-e^{-r_1t})\right]^{-n}+ns^ne^{-nr_1t}\left[1-s(1-e^{-r_1t})\right]^{-(n+1)}\left(1-e^{-r_1t}\right)\nonumber\\
					&=&\frac{ns^{n-1}e^{-nr_1t}}{\left[1-s(1-e^{-r_1t})\right]^{n}}\big[1+ \frac{s(1-e^{-r_1t})}{1-s(1-e^{-r_1t})}\big]\nonumber\\
					&=& \frac{ns^{n-1}e^{-nr_1t}}{\left[1-s(1-e^{-r_1t})\right]^{(n+1)}}\left[1-s(1-e^{-r_1t})+s(1-e^{-r_1t})\right]\nonumber\\
					&=& \frac{ns^{n-1}e^{-nr_1t}}{\left[1-s(1-e^{-r_1t})\right]^{(n+1)}}\label{rhs}, 
				\end{eqnarray}
				and the left hand-side is 
				\begin{eqnarray}
					\frac{\partial G}{\partial t}&=&-nr_1s^ne^{-nr_1t}\left[1-s(1-e^{-r_1t})\right]^{-n}+ns^{n+1}r_1e^{-(n+1)r_1t}\left[1-s(1-e^{-r_1t})\right]^{-(n+1)}\nonumber\\
					&=&\frac{nr_1s^ne^{-nr_1t}}{\left[1-s(1-e^{-r_1t})\right]^{n}}\left[\frac{se^{-r_1t}}{1-s(1-e^{-r_1t})}-1\right]\nonumber\\
					&=&\frac{nr_1s^ne^{-nr_1t}}{\left[1-s(1-e^{-r_1t})\right]^{(n+1)}}\left[se^{-r_1t}-1 +s-se^{-r_1t}\right]= \frac{nr_1s^ne^{-nr_1t}(s-1)}{\left[1-s(1-e^{-r_1t})\right]^{(n+1)}}\nonumber\\
					&=&r_1s(s-1)\overbrace{\left[\frac{ns^{n-1}e^{-nr_1t}}{\left[1-s(1-e^{-r_1t})\right]^{(n+1)}}\right]}^{\textit{this matches \eqref{rhs}}} = r_1s(s-1)\frac{\partial G}{\partial s}.
				\end{eqnarray}
			\end{proof}

			\item $k=2$:\\
			There are two amplification phases with rates $r_1$ and $r_2$, respectively. The first one runs from time $t=0$  to $t=\tau_1$, and the second one runs from $t=\tau_1$ to  $t=\tau_1+\tau_2$. In the second phase, the probability generating function takes exactly the same general functional form as in the first phase, albeit with a different initial condition. Specifically, we have
			\begin{equation}
				\nonumber
				G(s,t) =  \varPsi\left(\frac{s-1}{s}e^{r_2(t-\tau_1)}\right),
			\end{equation}
			with the initial condition (at time $t = \tau_1$)
			\begin{equation}
				\nonumber
				G(s,\tau_1) = \varPsi\left(\frac{s-1}{s}\right)
				= \frac{s^ne^{-nr_1\tau_1}}{\left[1-s\left(1-e^{-r_1\tau_1}\right)\right]^n}.
			\end{equation}
			Using the same procedure as in the case when $k=1$, we obtain 
			\begin{eqnarray}
				\label{pde22}
				G(s,t)& =& \varPsi\left(\frac{s-1}{s}e^{r_2(t-\tau_1)}\right) \nonumber \\
				&=& \frac{\left(\frac{1}{1-\frac{s-1}{s}e^{r_2(t-\tau_1)}}\right)^ne^{-nr_1\tau_1}}{\left[1- {\left(\frac{1}{1-\frac{s-1}{s}e^{r_2(t-\tau_1)}}\right)} \left(1-e^{-r_1\tau_1}\right)\right]^n}\nonumber\\
				&=&\frac{s^ne^{-n\left[r_2t+(r_1-r_2)\tau_1\right]}}{\left[1-s\left(1-e^{-\left[r_2t+(r_1-r_2)\tau_1\right]}\right)\right]^n}.
			\end{eqnarray}
			
			The right side of  \eqref{pde22} equals \eqref{pde} when $k=2$, as expected. 
			
			\item 
			We assume the statement is true for $t \in I_k$, that is
			$$ G(s,t)  = \left[\frac{se^{-z}}{1-s(1-e^{-z})}\right]^n, $$
			where $z = r_kt + \sum_{i=1}^{k-1} (r_i-r_{k})\tau_i$, and we prove it for $t \in I_{k+1}$.
			As before, in phase $k+1$, the generating function has the functional form
			$$ G(s,t) = \varPsi\left(\frac{s-1}{s}e^{r_{k+1}(t-\sum_{i=1}^{k}\tau_i)}\right).$$
			At time $t=\sum_{i=1}^{k}\tau_i$, by the induction step, we have
			$$ G(s,t) = \varPsi\left(\frac{s-1}{s}\right) =  \left[\frac{se^{-z}}{1-s(1-e^{-z})}\right]^n.$$ 
			Using the same arguments as before, we find that, for $t \in I_{k+1}$, 
			\begin{eqnarray}
				G(s,t) &=& \varPsi\left(\frac{s-1}{s}e^{r_{k+1}(t-\sum_{i=1}^{k}\tau_i)}\right)\nonumber \\				
				&=& \left[\frac{\frac{1}{1-\frac{s-1}{s}e^{r_{k+1}(t-\sum_{i=1}^{k}\tau_i)}}e^{-z}}{1-\frac{1}{1-\frac{s-1}{s}e^{r_{k+1}(t-\sum_{i=1}^{k}\tau_i)}}(1-e^{-z})}\right]^n 			\nonumber\\
				&=& \frac{s^ne^{-n\left[r_{k+1}t + \sum_{i=1}^{k} (r_i-r_{k})\tau_i\right]}}{\left[1-s\left(1-e^{-\left[r_{k+1}t + \sum_{i=1}^{k} (r_i-r_{k})\tau_i\right]}\right)\right]^n},
				\label{pdee}
			\end{eqnarray}			
			and this ends the proof of Theorem \ref{thm1}.  
		\end{itemize} 
	\end{proof}
	
	\subsubsection{Proof of Corollary \ref{thm2}}\label{pthm2}
	\begin{proof}
		From Theorem \ref{thm1}, we know that the generating function for $P(x|n,\vec{r},t,\vec{\tau})$ is given by
		$$ G(s,t) =\left[\frac{se^{-z}}{1-s(1-e^{-z})}\right]^n. $$
		Let $p = e^{-z}$ and $q = 1-p$. Then, 
		$$ G(s,t) = \left[\frac{sp}{1-sq}\right]^n  = (sp)^n\left[1-sq\right]^{-n}.
		$$ 
		$P(x|n,\vec{r},t,\vec{\tau})$ is the coefficient of $s^x$ in the power series expansion of $ G(s,t)$, given by
		\begin{eqnarray}
			G(s,t) &= &(sp)^n\left[1-sq\right]^{-n} = (sp)^n\sum_{i=0} \binom{-n}{i}(-sq)^i = (sp)^n\sum_{i=0} \binom{-n}{i}(-1)^i(sq)^i.\label{c1}
		\end{eqnarray}
		But
		\begin{eqnarray}
			\binom{-n}{i}& =& \frac{-n(-n-1)(-n-2)(-n-3)\ldots(-n-(i-2))(-n-(i-1))}{1.2.3.\ldots (i-1)i} \nonumber\\
			&=& (-1)^i\frac{n(n+1)(n+2)(n+3)\ldots (n+(i-2))(n+(i-1))}{i!}\nonumber\\
			&=& (-1)^i\binom{n+i-1}{i}\nonumber\\
			&\Longrightarrow &(-1)^i\binom{-n}{i} =\binom{n+i-1}{i}.\label{c2}
		\end{eqnarray}
		Substituting \eqref{c2} into \eqref{c1} gives
		\begin{eqnarray}
			G(s,t) &=& (sp)^n\sum_{i=0} \binom{-n}{i}(-1)^i(sq)^i = (sp)^n\sum_{i=0} \binom{n+i-1}{i}(sq)^i.
		\end{eqnarray}
		Let $x = n+i$. Then,
		\begin{eqnarray}
			G(s,t) &=& (sp)^n\sum_{x=n}^{\infty} \binom{x-1}{x-n}(sq)^{x-n} = \sum_{x=n}^{\infty}\binom{x-1}{x-n}p^nq^{x-n}s^x.
		\end{eqnarray}
		The probability of having $x$ molecules at time $t$, $P(x,t)$, is therefore given by 
		\begin{equation}
			P(x,t) = \binom{x-1}{x-n}p^nq^{x-n} = \binom{x-1}{n-1}e^{-nz}\left(1-e^{-z}\right)^{x-n}.
		\end{equation}
		
	\end{proof}
	
	\begin{corollary}\label{cor2}
		The probability distribution given in Theorem \ref{thm2} solves the Chapman-Kolmogorov equation given by  \eqref{me}.
	\end{corollary}
	\begin{proof}
		\begin{eqnarray}
			\frac{\partial P(x,t)}{\partial t} & = \binom{x-1}{x-n}\left[-nr_ke^{-z}\left(1-e^{-z}\right)^{x-n} + r_k(x-n)e^{-2z}\left(1-e^{-z}\right)^{x-n-1} \right]\nonumber\\
			& = r_k\binom{x-1}{x-n}e^{-z}\left(1-e^{-z}\right)^{x-n}\left[-n +(x-n)\frac{e^{-z}}{1-e^{-z}}\right]\nonumber\\
			& = 	r_k\binom{x-1}{x-n}e^{-z}\left(1-e^{-z}\right)^{x-n}\left[-n +(x-n)\frac{e^{-z}}{1-e^{-z}}-\frac{x-n}{1-e^{-z}}+\frac{x-n}{1-e^{-z}}\right]\nonumber\\
			&= 	r_k\binom{x-1}{x-n}e^{-z}\left(1-e^{-z}\right)^{x-n}\left[ \frac{x-n}{1-e^{-z}} -n  +\frac{(x-n)}{1-e^{-z}}\left(e^{-z}-1\right)\right]\nonumber\\
			&= 	r_k\binom{x-1}{x-n}e^{-z}\left(1-e^{-z}\right)^{x-n}\left[ \frac{x-n}{1-e^{-z}} -x\right]\nonumber \\
			&=	r_k\left[(x-n)\binom{x-1}{x-n}\right]e^{-z}\left(1-e^{-z}\right)^{x-n-1} - r_kx\binom{x-1}{x-n}e^{-z}\left(1-e^{-z}\right)^{x-n}\nonumber\\
            & =r_k(x-1)\left[\binom{x-2}{x-n-1}e^{-z}\left(1-e^{-z}\right)^{x-n-1}\right] -   r_kx\left[\binom{x-1}{x-n}e^{-z}\left(1-e^{-z}\right)^{x-n}\right]\nonumber\\
			&= r_k(x-1)P(x-1,t) -r_kxP(x,t),
        \label{coll}
		\end{eqnarray}
		which equals the right-hand side of \eqref{me}.
	\end{proof}

	\subsubsection{Proof of Theorem \ref{thm3}}\label{pthm3}
	\begin{proof} 
		We  prove Theorem \eqref{thm3} by mathematical induction on $k$.
		\begin{itemize}
			\item $k=1$: \\ There is only one phase with amplification efficiency $r_1$.
			Recall that the Chapman-Kolmogorov forward equation for the dynamics of $P\left(x,t\right)$ is given by \eqref{me}, with the initial condition
			$$P\left(x,0\right) = \frac {e^{-\lambda} \lambda^x} {x!}. $$
			Using the same arguments as above, we can write the generating function for $P\left(x,t\right)$ as
			\begin{equation}
				\label{erf2}
				G(s,t) = \varPsi\left( \frac{s-1}{s} e^{r_1t} \right),
			\end{equation}
			with the initial condition
			\begin{eqnarray}
				G(s,0) &=& \varPsi\left(\frac{s-1}{s}\right) = \sum_{x=0}^{\infty} s^xP(x,0) = e^{\lambda\left( s-1 \right)}.
			\end{eqnarray}
			Therefore, 
			\begin{eqnarray}
				\label{pde212}
				G(s,t) &=& \varPsi\left(\frac{s-1}{s} e^{r_1t}\right) \nonumber\\
				&=& e^{\lambda\left( \frac{1}{1 - \frac{s-1}{s} e^{r_1t}}-1 \right)} \nonumber\\
				&=& e^{\left(\frac{\lambda (s-1)}{1-s\left(1-e^{-r_1t}\right)}\right)}.
			\end{eqnarray}
			Equation \eqref{pde212} matches  \eqref{thm21} when $k=1$.
			
			\begin{corollary}\label{cor3}
				The generating function given by \eqref{pde212} solves Equation \eqref{pde}.
				
				\begin{proof}
					Differentiate the right hand side of \eqref{pde212} with respect to $s$. 
				\end{proof}
			\end{corollary}

			\item $k=2$:\\
			There are two phases with amplification efficiencies $r_1$ and $r_2$, respectively. The first phase runs from time $t=0$ to $t=\tau_1$, and the second one runs from $t=\tau_1$ to  $t=\tau_1+\tau_2$. As before, for $t \in I_2$ the probability generating function has the form
			\begin{equation}
				\nonumber
				G(s,t) =  \varPsi\left(\frac{s-1}{s}e^{r_2(t-\tau_1)}\right),
			\end{equation}
			with the initial condition (at time $t=t_1$)
			\begin{equation}
				\nonumber
				G(s,\tau_1) = \varPsi\left(\frac{s-1}{s}\right)
				= e^{\left[\frac{\lambda (s-1)}{1-s\left(1-e^{-r_1t}\right)}\right]}.
			\end{equation}
			Therefore,
			\begin{eqnarray}
				\label{pde222}
				G(s,t)& =& \varPsi\left(\frac{s-1}{s}e^{r_2(t-\tau_1)}\right) \nonumber \\
				&=&e^{\left[\frac{\lambda (\frac{1}{1 - \frac{s-1}{s}e^{r_2(t-\tau_1)}} - 1)}{1 - \frac{1}{1 - \frac{s-1}{s}e^{r_2(t-\tau_1)}}\left(1-e^{-r_1t}\right)}\right]}\nonumber\\
				&=& e^{\left[\frac{\lambda(s-1)}{1 -s\left(1-e^{-(r_2t+(r_1-r_2)\tau_1)}\right)}\right]}.
			\end{eqnarray}
			The right side of  \eqref{pde222} equals \eqref{thm21} for the case $k=2$. 
			
			\item We assume the statement is true for $t \in I_k$, that is
			$$ G(s,t)  = e^{\left[\frac{\lambda(s-1)}{1-s(1-e^{-z})}\right]}, $$
			where $z = r_kt + \sum_{i=1}^{k-1} (r_i-r_{k})\tau_i$, and we prove it for $t \in I_{k+1}$.
			
			In phase $k+1$, the probability generating function has the form
			$$ G(s,t) = \varPsi\left(\frac{s-1}{s}e^{r_{k+1}(t-\sum_{i=1}^{k}\tau_i)}\right).$$
			By the induction step, the initial condition (at time $t=\sum_{i=1}^{k}\tau_i$) is given by
			$$ G(s,t) = \varPsi\left(\frac{s-1}{s}\right) =  e^{\left[\frac{\lambda(s-1)}{1-s(1-e^{-z})}\right]}.$$ 
			Therefore, for $t \in I_{k+1}$, we have
			\begin{eqnarray}
				G(s,t) &=& \varPsi\left(\frac{s-1}{s}e^{r_{k+1}(t-\sum_{i=1}^{k}\tau_i)}\right) 	
				\nonumber\\
				& =& e^{\left[\frac{\lambda(\frac{1}{1 - \frac{s-1}{s}e^{r_{k+1}(t-\sum_{i=1}^{k}\tau_i)}} - 1)}{1 - \frac{1}{1-\frac{s-1}{s}e^{r_{k+1}(t-\sum_{i=1}^{k}\tau_i)}} (1-e^{-z})}\right]} \nonumber \\
				&=&  e^{\left[\frac{\lambda(s-1)}{1-s(1-e^{-z^\prime})}\right]},
				\label{pdee2}
			\end{eqnarray}
			where $z^\prime = r_{k+1}t + \sum_{i=1}^{k} (r_i-r_{k+1})\tau_i$,
			and this ends the proof of Theorem \ref{thm3}.  
		\end{itemize} 
	\end{proof}
	\subsubsection{Proof of Corollary \ref{thm4}}\label{pthm4}
	\begin{proof}
		Recall that $P(x|\lambda,\vec{r},t,\vec{\tau})$ is the coefficient of $s^x$ in the power series expansion of the probability generating function given in Theorem \ref{thm3}, that is 
		\begin{eqnarray}
			P(x|\lambda,\vec{r},t,\vec{\tau}) & =& \frac{1}{x!}\frac{\partial^{x}G(s,t)}{\partial s^x}\big|_{s=0}.
		\end{eqnarray}
		Now, let us compute the partial derivatives of $G(s,t)$ with respect to $s$. For brevity, we set $$a=e^{-z}, v=(1-e^{-z})=(1-a), u = \lambda e^{-z}= a\lambda, \;\;\text{and}\;\; Q(s,t)= 1-s(1-e^{-z}) = 1-sv.$$
		Observe that 
		$$ G(s,t)= e^{\frac{\lambda(s-1)}{Q}}, Q(0,t)= 1, G(0,t)= e^{-\lambda},  \frac{\partial Q(s,t)}{\partial s} =   \frac{\partial Q(s,t)}{\partial s}\big|_{s=0} = -v $$
		and 
		\begin{eqnarray}
			\frac{\partial }{\partial s}G(s,t) &=&u \frac{G(s,t)}{Q^2}\implies\frac{\partial }{\partial s}G(s,t)\big |_{s=0}= u e^{-\lambda}, \nonumber\\
			\frac{\partial^2 }{\partial s^2}G(s,t) &=& \frac{ue^{-\lambda}}{Q^4}\left[\frac{Q^2uG(s,t)}{Q^2}+2vQG(s,t)\right]=u\left[u+2vQ\right]\frac{G(s,t)}{Q^4}\nonumber\\
			&\implies &\frac{\partial^2 }{\partial s^2}G(s,t)\big |_{s=0}= e^{-\lambda}\left[u^2+2uv\right]\nonumber,\\
			\frac{\partial^3 }{\partial s^3}G(s,t) &=&\frac{u\left[u+2v\right]}{Q^8}\left[uG(s,t)(u+2vQ)Q^2-2v^2G(s,t)Q^4+4vG(s,t)(u+2vQ)Q^3\right]\nonumber\\
			&= &u\frac{G(s,t)}{Q^6}\left[u^2+6uvQ+6v^2Q^2\right]\nonumber\\
			&\implies&\frac{\partial^3 }{\partial s^3}G(s,t)\big|_{s=0}= e^{-\lambda}(u^3+6u^2v+6uv^2),\nonumber\\
			\frac{\partial^4 }{\partial s^4}G(s,t)&=&u\frac{G(s,t)}{Q^8}\left[u^3+12u^2vQ+36uv^2Q^2+24v^3Q^3\right]\nonumber\\
			&\implies& \frac{\partial^4 }{\partial s^4}G(s,t)\big|_{s=0}= e^{-\lambda}\left[u^4+12u^3v+36u^2v^2+24uv^3\right]\nonumber\\
			\frac{\partial^5 }{\partial s^5}G(s,t)&=&u\frac{G(s,t)}{Q^8}\left[u^4 +20u^3vQ+120u^2v^2Q^2+240uv^3Q^3+120v^4Q^4\right],\nonumber\\
			&\implies& \frac{\partial^4 }{\partial s^5}G(s,t)\big|_{s=0}= e^{-\lambda}\left[u^5 +20u^4v+120u^3v^2+240u^2v^3+120uv^4\right]\label{formula}.
		\end{eqnarray}
		By closely examining the coefficients of powers of the terms $u,v \;\text{and}\;uv$ in \eqref{formula}, the following combinatorial triangle emerges\newline
		
		\begin{center}
			\begin{tabular}{>{$}l<{$}|*{6}{c}}
				\multicolumn{1}{l}{$x$} &&&&&&\\\cline{1-1} 
				1 &1&&&&&\\
				2 &1&2&&&&\\
				3 &1&6&6&&&\\
				4 &1&12&36&24&&\\
				5 &1&20&120&240&120&\\\hline
				\multicolumn{1}{l}{} &1&2&3&4&5\\\cline{2-7}
				\multicolumn{1}{l}{} &\multicolumn{6}{c}{$i$}
			\end{tabular}
		\end{center}
		In particular, the $(x,i)$'th entry of the triangle is given by 
		\begin{equation}
			T(x,i)= \binom{x}{i-1}\binom{x-1}{i-1}(i-1)!, \;\;\text{for}\;\ i=1,2,\ldots x.
		\end{equation}
		Thus
		\begin{eqnarray}
			P(x|\lambda,\vec{r},t,\vec{\tau}) &=& \frac{1}{x!}\frac{\partial^x }{\partial s^x}G(s,t)\Big|_{s=0} =\frac{e^{-\lambda}}{x!}\sum_{i=1}^{x}T(x,i)\nonumber\\
			&=&\frac{ e^{-\lambda}}{x!}\sum_{i=1}^{x}\binom{x}{i-1}\binom{x-1}{i-1}(i-1)! \left(\lambda e^{-z}\right)^{x-i+1}\left(1-e^{-z}\right)^{i-1}\nonumber\\
			&=&e^{-\lambda}\sum_{i=1}^{x} \frac{1}{(x-i+1)!} \binom{x-1}{i-1}\left(\lambda e^{-z}\right)^{x-i+1}\left(1-e^{-z}\right)^{i-1}. \label{eq43}
		\end{eqnarray}
		Setting $k = x - i + 1$ in \eqref{eq43} gives the desired result.
	\end{proof}
	
	\begin{corollary}\label{cor4}
		The probability distribution given in Theorem \ref{thm4} solves \eqref{me}.
	\end{corollary}
	\begin{proof}
		Differentiate the right-hand-side of Equation \eqref{eq43} with respect to $t$. 
			\end{proof}
	
	We will now derive the probability density function (pdf), mean, variance, and cumulative density function (cdf) of the $Ct$ value. We will consider two different cases, namely:
  \begin{enumerate}
            \item when the initial state of the PCR process is deterministic, and the PCR phase lengths and amplification efficiencies are given; and
            \item when the initial state is Poisson-distributed, and the phase lengths and amplification efficiencies are given.
        \end{enumerate}

\subsubsection{Case 1: The initial state is deterministic, and the phase lengths and amplification efficiencies are given}\label{statsn}
		\begin{itemize}
\item \textbf{General form of the pdf}\\
	Consider the PCR process described in Theorem \ref{thm1}. The process begins with $n$ cDNA molecules, which are amplified across up to $p$ successive phases $\{I_i\}$ of lengths $\vec{\tau} = (\tau_1, \tau_2, \dots , \tau_p)$ at the corresponding amplification efficiencies $\vec{r} = (r_1, r_2, \ldots, r_p)$. By definition, the $Ct$ value $t$ is the time at which the number of molecules reaches the quantification threshold, which we denote by $x$. By Bayes' theorem, a general expression for the pdf of $t$ is given by 
	\begin{eqnarray}
		P(t|n,\vec{r}, \vec{\tau}, x) &=& \frac{P(n,\vec{r},  \vec{\tau}, x|t)P(t)}{P(n,\vec{r}, \vec{\tau}, x)}.
        \label{bayes1}
	\end{eqnarray}
	Since $n$ is independent of $\vec{r}$, $\vec{\tau}$, and $t$, and $\vec{r}$ is also independent of $t$ and of the values taken by the entries of $\vec{\tau}$, we re-write the numerator of the right-hand-side of \eqref{bayes1} as follows:
    \begin{eqnarray}
        P(n,\vec{r},  \vec{\tau}, x|t)P(t) &=& P(x|n,\vec{r},  \vec{\tau}, t)P(n,\vec{r},  \vec{\tau}|t)P(t)\nonumber\\
        &=& P(x|n,\vec{r},  \vec{\tau}, t)P(n|\vec{r},  \vec{\tau},t)P(\vec{r},  \vec{\tau}|t)P(t)\nonumber\\
        &=& P(x|n,\vec{r},  \vec{\tau}, t)P(n)P(\vec{r}|\vec{\tau},t)P(\vec{\tau}|t)P(t)\nonumber\\
        &=& P(x|n,\vec{r},  \vec{\tau}, t)P(n)P(\vec{r})P(t|\vec{\tau})P(\vec{\tau}).
	\end{eqnarray}
    Similarly, the denominator of the right-hand-side of \eqref{bayes1} can be simplified to
    \begin{eqnarray}
         P(n,\vec{r},  \vec{\tau}, x) &=& \int_{\sum_{i=1}^{k-1}\tau_i}^{\infty} P(n,\vec{r},  \vec{\tau}, x|t)P(t) dt\nonumber\\
         &=& P(n)P(\vec{r})P(\vec{\tau}) \int_{\sum_{i=1}^{k-1}\tau_i}^{\infty} P(x|n,\vec{r},  \vec{\tau}, t)P(t|\vec{\tau}) dt.
    \end{eqnarray}
    Therefore, \eqref{bayes1} can be re-written as
    \begin{eqnarray}
        P(t|n,\vec{r}, \vec{\tau}, x) &=& \frac{P(x|n,\vec{r},  \vec{\tau}, t)P(t|\vec{\tau})}{ \int_{\sum_{i=1}^{k-1}\tau_i}^{\infty} P(x|n,\vec{r},  \vec{\tau}, t)P(t|\vec{\tau}) dt}.
		\label{genarl}
	\end{eqnarray}
	We will derive the pdf, mean, variance, and cdf of the $Ct$ value for a PCR process with an arbitrary number of phases $p$. Without loss of generality, we suppose that $t \in I_k, k \leq p$. We will assume a uniform prior density for $t$ given $\vec{\tau}$. As we will demonstrate later, this assumption produces very similar results to those we obtain by assuming a Jeffreys prior \citep{jeffreys}. We will state results for the case of a single-phase PCR process whenever these cannot be readily gleaned from the general results.

    We will first consider the case when the lengths of the intermediate phases, recorded in the vector $\vec{\tau}$, are given. This is useful, for example, when it is of interest to estimate the lengths of such phases from data. We will then show how to marginalize $\vec{\tau}$ out of the pdf. 
		\item \textbf{pdf}\\
		Using the posterior density given in Equation \eqref{genarl} and the likelihood function given in Corollary \ref{thm2}, we obtain the following functional form for the pdf:
		\begin{eqnarray}
			P(t|n,\vec{r},\vec{\tau},x) &\propto& e^{-nz}\left(1-e^{-z}\right)^{x-n}, 
			\label{post21}
		\end{eqnarray}
where
\begin{equation}
z= r_kt + \sum_{i=1}^{k-1}(r_i-r_k)\tau_i,
\label{z_eqn}
\end{equation}
$\vec{r}=(r_1,r_2,...,r_k)$ is a vector of amplification efficiencies, $\vec{\tau} = (\tau_1,\tau_2,...,\tau_k)$ is a vector of phase lengths, and we have used a uniform prior for $t$.

The normalizing constant is given by 
		\begin{equation}
			C = \int_{\sum_{i=1}^{k-1}\tau_i}^{\infty} e^{-nz}(1-e^{-z})^{x-n} dt \label{nc2}.
		\end{equation}
		Let $ w = e^{-z}.$ Then,
		\begin{eqnarray}
			C &=& \frac{1}{r_k} \int_{0}^{\theta} w^{n-1} (1-w)^{x-n} dw\nonumber\\
			&=& \frac{B_{\theta}(n,x-n+1)}{r_k},
		\end{eqnarray}
   where
  \begin{equation}
      \theta = e^{-\sum_{i=1}^{k-1}r_i\tau_i}. \label{theta_eqn}
  \end{equation}
		Therefore, the pdf is given by 
		\begin{eqnarray}
		P(t|n,\vec{r},\vec{\tau},x) &=& \frac{r_k e^{-nz}\left(1-e^{-z}\right)^{x-n}}{B_{\theta}(n,x-n+1)}.
			\label{pdf1}
		\end{eqnarray}
  For the single-phase process, $\theta = 1$, so the pdf simplifies to
 	\begin{eqnarray}
		P(t|n,r_1,x) &=& \frac{r_1 e^{-nr_1t}\left(1-e^{-r_1t}\right)^{x-n}}{B(n,x-n+1)}.
			\label{pdf11}
		\end{eqnarray}
        Note that in some cases (eg. when knowledge of the lengths of individual PCR amplification phases is not of interest), it may be useful to marginalize $\vec{\tau}$ out of $P(t|n,\vec{r},\vec{\tau},x)$. This can be achieved by using the fact that
        \begin{eqnarray}
		P(t|n,\vec{r},x) &=& \int_{\Omega}P(t,\vec{\tau}|n,\vec{r},x) d \vec{\tau}\nonumber\\
        &=& \int_{\Omega}P(t|n,\vec{r},\vec{\tau},x) P(\vec{\tau}|n,\vec{r},x) d \vec{\tau},
			\label{pdf10}
		\end{eqnarray}
        where $\Omega$ is the domain of $\vec{\tau}$.
        
		In addition, note that an alternative formulation of the prior for $t$, based on an approach proposed by Jeffreys \citep{jeffreys} for generating priors that are invariant to reparametrization, is the following:
		\begin{equation}
		p(t|\vec{\tau}) \propto \sqrt{|I(t|\vec{\tau})|},
		\end{equation}
		where $I(t|\vec{\tau})$ is the Fisher information of the likelihood function and is given by
		\begin{eqnarray}
		I(t|\vec{\tau}) &=& \mathbb{E}_X \Big[\left(\frac{\partial}{\partial t} \ln P(x|n,\vec{r},\vec{\tau},x) \right)^2 \Big]\nonumber\\
		&=& \mathbb{E}_X \Big[ \frac{r_k^2 \left(w^2x^2 - 2nwx + n^2 \right)}{(1-w)^2} \Big]\nonumber\\
        &=& \frac{r_k^2 w^2}{(1-w)^2} \mathbb{E}_X \Big[ x^2 \Big] - \frac{2nr_k^2 w}{(1-w)^2} \mathbb{E}_X \Big[ x \Big] + \frac{n^2 r_k^2 }{(1-w)^2}\nonumber\\
        &=& \frac{r_k^2}{(1-w)^2} \Bigg[ w^2 \sum_{j=1}^{\infty} x^2 \binom{x-1}{j-1}w^n(1-w)^{x-n} - 2nw \sum_{j=1}^{\infty} x \binom{x-1}{j-1}w^n(1-w)^{x-n} + n^2 \Bigg],\nonumber\\
        \label{fisher1}
		\end{eqnarray}
		where $w = e^{-z}$.
        
        Observe that
        \begin{eqnarray}
		\sum_{x=1}^{\infty} x \binom{x-1}{n-1}w^n(1-w)^{x-n} &=& \sum_{x=1}^{\infty} \frac{x!}{(x-n)!(n-1)!} w^n(1-w)^{x-n}\nonumber\\
        &=& \sum_{y=2}^{\infty} \frac{(y-1)!}{(y-m)!(m-2)!} w^{m-1}(1-w)^{y-m}\nonumber\\
        &=& \frac{m-1}{w} \sum_{y=1}^{\infty} \frac{(y-1)!}{(y-m)!(m-1)!} w^{m}(1-w)^{y-m}\nonumber\\
        &=& \frac{m-1}{w} \nonumber\\
        &=& \frac{n}{w}
        \label{temp1}
		\end{eqnarray}
        and
        \begin{eqnarray}
		\sum_{x=1}^{\infty} x^2 \binom{x-1}{n-1}w^n(1-w)^{x-n} &=& \sum_{x=1}^{\infty} \frac{x!x}{(x-n)!(n-1)!} w^n(1-w)^{x-n}\nonumber\\
        &=& \sum_{y=2}^{\infty} \frac{(y-1)!(y-1)}{(y-m)!(m-2)!} w^{m-1}(1-w)^{y-m}\nonumber\\
        &=& \sum_{y=1}^{\infty} \frac{(y-1)!y}{(y-m)!(m-2)!} w^{m-1}(1-w)^{y-m} - \frac{m-1}{w} \nonumber\\
         &=& \sum_{y'=2}^{\infty} \frac{(y'-1)!}{(y'-m')!(m'-3)!} w^{m'-2}(1-w)^{y'-m'} - \frac{m-1}{w} \nonumber\\
        &=& \frac{(m'-1)(m'-2)}{w^2} \sum_{y'=1}^{\infty} \frac{(y'-1)!}{(y'-m')!(m'-1)!} w^{m'}(1-w)^{y'-m'} - \frac{m-1}{w} \nonumber\\
        &=& \frac{(m'-1)(m'-2)}{w^2} - \frac{m-1}{w} \nonumber\\
        &=& \frac{n(n+1)}{w^2} - \frac{n}{w} \;, \nonumber\\
        \label{temp2}
		\end{eqnarray}
        where $m=n+1, m' = m+1, y=x+1, y'=y+1$.
		
        Plugging \eqref{temp1} and \eqref{temp2} into \eqref{fisher1}, we obtain 
		\begin{eqnarray}
		I(t|\vec{\tau}) &=& \frac{n r_k^2}{1-w} \nonumber\\
		&\implies& p(t|\vec{\tau}) \propto \frac{1}{\sqrt{1-w}} = \frac{1}{\sqrt{1-e^{-z}}}.
		\end{eqnarray}
		Using this prior, and following the steps we used earlier to derive \eqref{pdf1}, we find that the pdf is given by
		\begin{eqnarray}
        P(t|n,\vec{r},\vec{\tau},x) &\propto& e^{-nz}\left(1-e^{-z}\right)^{x-n-1/2} \nonumber\\
		\implies P(t|n,\vec{r},\vec{\tau},x) &=& \frac{r_k e^{-nz}\left(1-e^{-z}\right)^{x-n-1/2}}{B_{\theta}(n,x-n+1/2)},
		\end{eqnarray}
		which has a similar form as \eqref{pdf1}.
  
  For simplicity, we will continue to use a uniform prior for $t$.
		
		\item \textbf{Mean}\\
		The mean $Ct$ value is given by
		\begin{eqnarray}
			\mathbb{E}(t) &=& \frac{r_k \int_{\sum_{i}^{k-1}\tau_i}^{\infty} t e^{-nz} (1 - e^{-z})^{x-n} dt}{B_{\theta}(n,x-n+1)},\label{mean1c}
		\end{eqnarray}
		where $z$ is given by \eqref{z_eqn}.
  
		Let $w = e^{-z}. $
		Then,
		\begin{eqnarray}
			\mathbb{E}(t) &=& \frac{r_k \int_{\theta}^{0} \Big[ -\frac{(\ln w - \ln \theta')}{r_k} \Big] w^n(1-w)^{x-n} \Big(-\frac{dw}{r_kw} \Big)}{B_{\theta}(n,x-n+1)} \nonumber\\
			&=& \frac{\int_{\theta}^{0} \left(\ln w - \ln \theta' \right) w^{n-1}(1-w)^{x-n} dw}{r_k B_{\theta}(n,x-n+1)} \nonumber\\
			&=& \frac{ \ln \theta' \int_{0}^{\theta} w^{n-1}(1-w)^{x-n} dw -  \int_{0}^{\theta} \ln w \: w^{n-1}(1-w)^{x-n} dw }{r_k B_{\theta}(n,x-n+1)} \nonumber\\
			&=& \frac{\ln \theta'}{r_k} -  \frac{ \Big(\frac{\partial}{\partial n} + \frac{\partial}{\partial x} \Big) B_{\theta}(n,x-n+1) }{r_k B_{\theta}(n,x-n+1)} \nonumber\\
			&=& \frac{\ln \frac{\theta'}{\theta}}{r_k} + \frac{\Gamma(n)^2 {\theta}^n \: {}_3\tilde{F}_2(n,n,n-x;n+1,n+1;\theta)}{r_k B_{\theta}(n,x-n+1)}\nonumber\\
            &=& \sum_{i=1}^{k-1}\tau_i + \frac{\Gamma(n)^2 {\theta}^n \: {}_3\tilde{F}_2(n,n,n-x;n+1,n+1;\theta)}{r_k B_{\theta}(n,x-n+1)},
			\label{mean1}
		\end{eqnarray}
where $\theta$ is given by \eqref{theta_eqn}, $\psi(\cdot)$ is the first polygamma function (also called the digamma function), and
  \begin{equation}
      \theta' = \theta e^{r_k\sum_{i=1}^{k-1}\tau_i}. \label{thetap_eqn}
  \end{equation}
Note that for the single-phase process, $\theta = \theta' = 1$. In this case, using
$$ {}_3\tilde{F}_2(n,n,n-x;n+1,n+1;1) = \frac{n \Gamma(x-n+1)\left[\psi(x+1) - \psi(n) \right]}{\Gamma(n+1) \Gamma(x+1)}, $$
we find that the mean $Ct$ value is given by
\begin{equation}
\mathbb{E}(t) = \frac{\psi(x+1) - \psi(n)}{r_1}.
\label{mean11}
\end{equation} 

  \item \textbf{Variance}\\ 
		The variance of the $Ct$ value is given by $\mathbb{E}(t^2)- \mathbb{E}(t)^2 $, where
		\begin{eqnarray}
			\mathbb{E}(t^2) &=& \frac{r_k \int_{\sum_{i=1}^{k-1}\tau_i}^{\infty} t^2 e^{-nz} (1 - e^{-z})^{x-n} dt}{B_{\theta}(n,x-n+1)}, \nonumber\\\label{var1c}
		\end{eqnarray}
  and $z$ is given by \eqref{z_eqn}.
  
		Let $w = e^{-z}.$ Then,
		\begin{eqnarray}
			\mathbb{E}(t^2) &=& \frac{r_k \int_{\theta}^{0} \Big[ \frac{\ln w - \ln \theta'}{r_k} \Big]^2 w^n(1-w)^{x-n} \Big(-\frac{dw}{r_kw} \Big)}{B_{\theta}(n,x-n+1)}  \nonumber\\
			&=& \frac{\int_{0}^{\theta} \left(\ln w - \ln \theta' \right)^2  w^{n-1}(1-w)^{x-n} dw}{{r_k}^2  B_{\theta}(n,x-n+1) }  \nonumber\\
			&=& \frac{\int_{0}^{\theta} \left(\ln w \right)^2 w^{n-1}(1-w)^{x-n} dw - 2 \ln \theta' \int_{0}^{\theta} \ln w \: w^{n-1}(1-w)^{x-n} dw }{{r_k}^2 B_{\theta}(n,x-n+1) } + \nonumber\\
            && \frac{(\ln \theta')^2 \int_{0}^{\theta} \: w^{n-1}(1-w)^{x-n} dw }{{r_k}^2 B_{\theta}(n,x-n+1) }\nonumber\\
			&=& \frac{\Big(\frac{\partial ^2}{\partial n^2} + 2\frac{\partial ^2}{\partial n \partial x} + \frac{\partial ^2}{\partial x^2} - 2 \ln \theta' \Big(\frac{\partial}{\partial n} + \frac{\partial}{\partial x} \Big) \Big) B_{\theta}(n,x-n+1)}{{r_k}^2 B_{\theta}(n,x-n+1)} + \frac{(\ln\theta')^2 }{{r_k}^2 }  \nonumber\\
			&=& \frac{\Big(\frac{\partial ^2}{\partial n^2} + 2\frac{\partial ^2}{\partial n \partial x} + \frac{\partial ^2}{\partial x^2} \Big) B_{\theta}(n,x-n+1)}{{r_k}^2 B_{\theta}(n,x-n+1)} + \frac{2\ln\theta'\Gamma(n)^2 {\theta}^n \: {}_3\tilde{F}_2(n,n,n-x;n+1,n+1;\theta)}{r_k^2 B_{\theta}(n,x-n+1)}+\nonumber\\ 
            && \frac{\ln\theta' \ln\frac{\theta'}{{\theta}^2}}{r_k^2}\nonumber\\
            &=& \frac{\Big(\frac{\partial ^2}{\partial n^2} + 2\frac{\partial ^2}{\partial n \partial x} + \frac{\partial ^2}{\partial x^2} \Big) B_{\theta}(n,x-n+1)}{{r_k}^2 B_{\theta}(n,x-n+1)} + \frac{2\ln\theta'\Gamma(n)^2 {\theta}^n \: {}_3\tilde{F}_2(n,n,n-x;n+1,n+1;\theta)}{r_k^2 B_{\theta}(n,x-n+1)}+\nonumber\\ 
            && \left[\sum_{i=1}^{k-1}\tau_i \right]^2 - \left[\sum_{i=1}^{k-1} \frac{r_i\tau_i}{r_k} \right]^2,
			\label{var1}
		\end{eqnarray}
  		where $\theta$ is given by \eqref{theta_eqn} and $\theta'$ is given by \eqref{thetap_eqn}.

For the single-phase process, the second moment of the $Ct$ value is given by
\begin{eqnarray}
    \mathbb{E}(t^2) &=& \frac{\Big(\frac{\partial ^2}{\partial n^2} + 2\frac{\partial ^2}{\partial n \partial x} + \frac{\partial ^2}{\partial x^2} \Big) B(n,x-n+1)}{{r_k}^2 B(n,x-n+1)}\nonumber\\
    &=& \frac{\psi_1(n)-\psi_1(x+1) + \left[\psi(x+1) - \psi(n)\right]^2}{{r_k}^2},
\end{eqnarray}
where $\psi_1(\cdot)$ is the second polygamma function (also called the trigamma
function). Therefore, the variance is
\begin{equation}
    \textbf{Var}(t) = \frac{\psi_1(n)-\psi_1(x+1)}{{r_k}^2}.
\end{equation}	
		\item \textbf{cdf}\\
		The cdf of the $Ct$ value is given by
		\begin{eqnarray}
			F(t|n,\vec{r},\vec{\tau},x) & =&  \frac{r_k}{B_{\theta}(n,x-n+1)} \int_{\sum_{i=1}^{k-1}\tau_i}^{t} e^{-nz'} (1 - e^{-z'})^{x-n} ds,
		\end{eqnarray}
where $z' = r_ks + \sum_{i=1}^{k-1}(r_i-r_k)\tau_i$
		
		Let $w = e^{-z'}. $ Then,
		\begin{eqnarray}
			F(t|n,\vec{r},\vec{\tau},x) & =& \frac{\int_{e^{-z}}^{\theta} w^{n-1} (1-w)^{x-n} dw }{B_{\theta}(n,x-n+1)} \nonumber\\
			&=& \frac{B_{\theta}(n,x-n+1) - B_{e^{-z}}(n,x-n+1)}{B_{\theta}(n,x-n+1)}\nonumber\\
            &=& 1 - \frac{B_{e^{-z}}(n,x-n+1)}{B_{\theta}(n,x-n+1)},
			\label{cdf1}
		\end{eqnarray}
		where $\theta$ is given by \ref{theta_eqn}.

For the single-phase process, the cdf is given by
\begin{eqnarray}
F(t|n,r_1,x) &=& 1 - I_{e^{-r_1t}}(n,x-n+1),
\label{cdf11}
\end{eqnarray}
where $I_{e^{-r_1t}} (n, x-n+1) = \frac{B_{e^{-r_t}}(n,x-n+1)}{B(n,x-n+1)}$ is the regularized incomplete Beta function. Because the cdf is in closed analytical form,  we can apply the efficient inverse transform method to generate random samples of $Ct$ values as follows: 
	\begin{eqnarray}
		t &=& -\frac{\ln I_{1-u}^{-1}(n, x-n+1)}{r_1},\label{sampling}
	\end{eqnarray}
	where  $u$ is a real number sampled uniformly at random from  the interval $(0,1)$ and  $I^{-1}_{1-u}$ is the inverse of the regularized incomplete Beta function. To find a $Ct$ value that corresponds to a quantile  $q \in (0,1)$, simply replace $u$ in Equation \eqref{sampling} by $q$. 	
	
\item \textbf{Probability distribution of $n$}\\
We conclude by deriving the probability distribution of $n$, denoted $P(n|r_1,t,x)$, for the single-phase PCR process. This distribution can be used to estimate $n$ from measured $Ct$ values. It can also be used to calculate the LoD and LoQ of a PCR assay, as we demonstrated in the main text. The steps described below can also be used to derive $P(n|\vec{r},t,\vec{\tau})$, for a PCR process with an arbitrary number of phases, although this will not yield a closed-form result like we will obtain in the single-phase case.

By Bayes' Theorem, we have
\begin{eqnarray}
P(n|r_1,t,x) &\propto & \frac{w^{n}(1-w)^{x-n}}{B(n,x-n+1)},
\end{eqnarray} 
where
\begin{equation}
    w = e^{-r_1t}. \label{w_eqn}
\end{equation}

The normalizing constant is given by
\begin{eqnarray}
C &=& \sum_{n=1}^{x} \frac{w^n(1-w)^{x-n}}{B(n,x-n+1)}\nonumber\\
&=& \sum_{n=1}^{x} \frac{x! \: w^n(1-w)^{x-n}}{(n-1)! \: (x-n)!}.
\end{eqnarray} 
Let $m = n-1$. Then,
\begin{eqnarray}
C &=& \sum_{m=0}^{x-1} \frac{x! \: w^{m+1}(1-w)^{x-1-m}}{m! \: (x-1-m)!} \nonumber\\
   &=& xw \sum_{m=0}^{x-1} \binom{x-1}{m} \: w^{m}(1-w)^{x-1-m} \nonumber\\
   &=& xw.
\end{eqnarray} 
Therefore, we have
\begin{eqnarray}
P\left(n|r_1,t,x\right) &=& \frac{w^{n-1}(1-w)^{x-n}}{x B(n,x-n+1)}.
\label{pdfn}
\end{eqnarray}
Suppose that $t$ is a $Ct$ value generated by a PCR process with $n$ input molecules. If we replace $n$ by $\hat{n}$ in Equation \eqref{pdfn}, then the equation will give the probability that $\hat{n}$ will be obtained as the estimate of $n$ based on the data $t$. It is useful -- eg. for the purpose of determining the LoQ -- to calculate the probability that $\hat{n}$ will be obtained as the estimate of $n$ based on any data $t$ that can be generated by a PCR process with $n$ input molecules. This probability is given by
\begin{eqnarray}
P\left(\hat{n}|n,r_1,x\right) &=& \int_{0}^{\infty}P\left(\hat{n},t|n,r_1,x\right)dt\nonumber\\
&=& \int_{0}^{\infty}P\left(\hat{n}|n,r_1,t,x\right)P\left(t|n,r_1,x\right)dt\nonumber\\
&=& \int_{0}^{\infty}P\left(\hat{n}|r_1,t,x\right)P\left(t|n,r_1,x\right)dt\nonumber\\
&=& r_1 \int_{0}^{\infty} \frac{w^{\hat{n}-1}(1-w)^{x-\hat{n}}}{x B(\hat{n},x-\hat{n}+1)} \frac{w^{n}(1-w)^{x-n}}{B(n,x-n+1)}dt \nonumber\\
&=& \int_{0}^{1} \frac{w^{\hat{n}+n-2}(1-w)^{2x-m-n}}{x B(\hat{n},x-\hat{n}+1)B(n,x-n+1)}dw \nonumber\\
&=& \frac{B(\hat{n}+n-1,2x-\hat{n}-n+1)}{x B(\hat{n},x-\hat{n}+1)B(n,x-n+1)} = P(\hat{n}|n,x),
\label{like}
\end{eqnarray}
where, using \eqref{pdfn}, we have assumed that $\hat{n}$ is conditionally independent of $n$ given $t$.
Strikingly, \eqref{like} does not depend on $r_1$.
	\end{itemize}

 \subsubsection{Case 2: The initial state is Poisson-distributed, and the phase lengths and amplification efficiencies are given}\label{statslam}
	\begin{itemize}
\item \textbf{General form of the pdf}\\
	Let $t$ be the $Ct$ value of the PCR process described in Theorem \ref{thm3}. The process begins with a Poisson-distributed number of input DNA molecules, with mean $\lambda$, which are replicated across up to $p$ distinct phases with lengths $\vec{\tau} = (\tau_1, \tau_2, \dots , \tau_p)$ and amplification efficiencies $\vec{r} = (r_1, r_2, \ldots, r_p)$. As noted earlier, $t$ is the time at which the number of molecules reaches the quantification threshold, which we denote by $x$. Let us denote the pdf of $t$ by $P(t|\lambda,\vec{r}, \vec{\tau}, x)$. By Bayes' theorem, we have 
	\begin{eqnarray}
		P(t|\lambda, \vec{r}, \vec{\tau}, x) &=& \frac{P(\lambda, \vec{r}, \vec{\tau}, x|t)P(t)}{P(\lambda,\vec{r}, \vec{\tau}, x)}
	\end{eqnarray}
	However, $\lambda$ is independent of $\vec{r}$, $\vec{\tau}$, and $t$, while $\vec{r}$ is also independent of $t$ and of the precise values taken by the entries of $\vec{\tau}$. Therefore, by following the same steps we used earlier to derive \eqref{genarl}, we find that
	\begin{eqnarray}
		P(t|\lambda, \vec{r}, \vec{\tau}, x) &=&\frac{P(x|\lambda,\vec{r},t,\vec{\tau})P(t|\vec{\tau})}{\int_{\sum_{i=1}^{k-1}\tau_i}^{\infty} P(x|\lambda,\vec{r},t,\vec{\tau})P(t|\vec{\tau}) dt}
    \label{general2}.
	\end{eqnarray}
	
	We will derive the pdf, mean, variance, and cdf by assuming, without loss of generality, that $t \in I_k$, and then we will specify the functional forms taken by the results in the instructive case when $t \in I_1$. As before, for simplicity, we will use a uniform prior density for $t$. 
	
			\item \textbf{pdf}\\
		We derive the pdf of the $Ct$ value $t$ by using the general expression given in Equation \eqref{general2}, with the probability distribution of the number of molecules given in Theorem \ref{thm4} serving as the likelihood. Specifically,
		\begin{eqnarray}
			\label{post222}
			P(t|\lambda, \vec{r}, \vec{\tau},x) &\propto&  (1 - e^{-z})^x \sum_{j=1}^{x} \frac{\binom{x-1}{j-1}}{j!} \left( \frac{\lambda e^{-z}}{1 - e^{-z}} \right)^j \\
			&=& (1 - e^{-z})^x \sum_{j=0}^{x-1} \frac{\binom{x-1}{j}}{(j+1)!} \left( \frac{\lambda e^{-z}}{1 - e^{-z}} \right)^{j+1} \\
			&& \stackrel{\text{since } \binom{x}{j} = 0 \text{ for } j > x }{=} (1 - e^{-z})^x \sum_{j=0}^{\infty} \frac{\binom{x-1}{j}}{(j+1)!} \left( \frac{\lambda e^{-z}}{1 - e^{-z}} \right)^{j+1} \\
			&=& \lambda e^{-z}(1 - e^{-z})^{x-1} \sum_{j=0}^{\infty} \frac{(x-1)(x-2) \cdots (x-j)}{(j+1)! \: j!} \left( \frac{\lambda e^{-z}}{1 - e^{-z}} \right)^{j} \\
			&=& \lambda e^{-z}(1 - e^{-z})^{x-1} \sum_{j=0}^{\infty} \frac{(1-x)(2-x) \cdots (j-x)}{(j+1)! \: j!} \left( \frac{-\lambda e^{-z}}{1 - e^{-z}} \right)^{j} \\
			&=& \lambda e^{-z}(1 - e^{-z})^{x-1} \sum_{j=0}^{\infty} \frac{(1-x)_j}{(2)_j } \frac{\left( \frac{-\lambda e^{-z}}{1 - e^{-z}} \right)^{j}}{j!} \\
			&=& \lambda e^{-z}(1 - e^{-z})^{x-1} \: {}_1F_1 \left(1-x;2; \frac{-\lambda e^{-z}}{1 - e^{-z}} \right),
		\end{eqnarray}
		where $z$ is given by \eqref{z_eqn}, ${}_1F_1$ is the hypergeometric function (also called the Kummer confluent hypergeometric function of the first kind), defined as 
	\begin{equation}
		{}_1F_1\big(1-x;2;\frac{-\lambda e^{-r_1t}}{1-e^{-r_1t}}\big)= \sum_{j=0}^{\infty}\frac{(1-x)_j}{(2)_j}\frac{\left[\frac{-\lambda e^{-r_1t}}{1-e^{-r_1t}}\right]^j}{j!},\nonumber
	\end{equation}
and $(\alpha)_j$ denotes the rising factorial, i.e. $(\alpha)_j = \alpha(\alpha+1)(\alpha +2)\ldots(\alpha+j-1)$ with $(\alpha)_0 =1$.

		The normalizing constant is given by 
		\begin{eqnarray}
			C &=& \sum_{j=1}^{x} \frac{\binom{x-1}{j-1} \lambda^j}{j!} \int_{\sum_{i=1}^{k-1}\tau_i}^{\infty} e^{-jz} (1 - e^{-z})^{x-j} dt.
					\end{eqnarray}	
	Let $w = e^{-z}$. Then,
		\begin{eqnarray}
			C &=& \frac{1}{r_k} \sum_{j=1}^{x} \frac{\binom{x-1}{j-1} \lambda ^j}{j!} \int_{0}^{\theta} w^{j-1} (1-w)^{x-j} dw \nonumber\\
			&=& \frac{1}{r_k} \sum_{j=1}^{x} \frac{\binom{x-1}{j-1} \lambda ^j}{j!} B_{\theta}(j,x-j+1),
		\end{eqnarray}	
  where $\theta$ is given by \eqref{theta_eqn}.
	
 Therefore, the pdf is given by
		\begin{eqnarray}
			P(t|\lambda, \vec{r}, \vec{\tau},x) &=& \frac{r_k (1 - e^{-z})^x \sum_{j=1}^{x} \frac{\binom{x-1}{j-1}}{j!} \left( \frac{\lambda e^{-z}}{1 - e^{-z}} \right)^j}{\sum_{j=1}^{x} \frac{\binom{x-1}{j-1} \lambda ^j}{j!} B_{\theta}(j,x-j+1)} \nonumber\\
            &=& \frac{r_k \lambda e^{-z}(1 - e^{-z})^{x-1} \: {}_1F_1 \left(1-x,2, \frac{-\lambda e^{-z}}{1 - e^{-z}} \right)} {\sum_{j=1}^{x} \frac{\binom{x-1}{j-1} \lambda ^j}{j!} B_{\theta}(j,x-j+1)},
			\label{pdf2}
		\end{eqnarray}
		Recall that for the single-phase process, $\theta = 1$, so we have 
		\begin{eqnarray}
			 \sum_{j=1}^{x} \frac{\binom{x-1}{j-1} \lambda ^j}{j!} B(j,x-j+1) &=& \sum_{j=1}^{\infty} \frac{\binom{x-1}{j-1} \lambda ^j}{j!} B(j,x-j+1)\nonumber\\
            &=& \sum_{j=0}^{\infty} \frac{\binom{x-1}{j} \lambda ^{j+1}}{(j+1)!} B(j+1,x-j) \nonumber\\
		      &=&  \sum_{j=0}^{\infty} \frac{\lambda ^{j+1}}{(j+1)!}\nonumber\\
		      &=& \frac{e^{\lambda} - 1}{ x}, \label{simplification}
		\end{eqnarray}	
implying that the pdf is given by
\begin{eqnarray}
			P(t|\lambda, r_1,x) &=& \frac{r_1 x \lambda e^{-r_1t} (1 - e^{-r_1t})^{x-1} \: {}_1F_1 \left(1-x,2, \frac{-\lambda e^{-r_1t}}{1 - e^{-r_1t}} \right)} {e^{\lambda} - 1}.
			\label{pdf21}
		\end{eqnarray}
  Note that in some cases (eg. when knowledge of the lengths of individual PCR amplification phases is not of interest), it may be useful to marginalize $\vec{\tau}$ out of $P(t|\lambda,\vec{r},\vec{\tau},x)$. This can be achieved by using the fact that
        \begin{eqnarray}
		P(t|\lambda,\vec{r},x) &=& \int_{\Omega}P(t,\vec{\tau}|\lambda,\vec{r},x) d \vec{\tau}\nonumber\\
        &=& \int_{\Omega}P(t|\lambda,\vec{r},\vec{\tau},x) P(\vec{\tau}|\lambda,\vec{r},x) d \vec{\tau},
			\label{pdf20}
		\end{eqnarray}
  where $\Omega$ is the domain of $\vec{\tau}$.
\item \textbf{Mean} \\
The mean $Ct$ value is given by
		\begin{eqnarray}
			\mathbb{E}(t)&=& \frac{r_k}{\sum_{j=1}^{x} \frac{\binom{x-1}{j-1} \lambda ^j}{j!} B_{\theta}(j,x-j+1)} \sum_{j=1}^{x} \frac{\binom{x-1}{j-1} \lambda ^j}{j!} \overbrace{ \int_{\sum_{i=1}^{k-1}\tau_i}^{\infty} t e^{-jz} (1 - e^{-z})^{x-j} dt}^{(D)}, \nonumber\\
		\end{eqnarray}
  where $z$ is given by \eqref{z_eqn}.
  
		Let $w = e^{-z}. $
		Then,
		\begin{eqnarray}
			D &\stackrel{\text{see \eqref{mean1}}}{=}& \frac{B_{\theta}(j,x-j+1) \sum_{i=1}^{k-1}\tau_i}{r_k} + \frac{\Gamma(j)^2 {\theta}^j \: {}_3\tilde{F}_2(j,j,j-x;j+1,j+1;\theta)}{r_k^2}.
			\label{mean21c}
		\end{eqnarray}	
Therefore
		\begin{eqnarray}
			\mathbb{E}(t)&=& \frac{\sum_{j=1}^{x} \frac{\binom{x-1}{j-1} \lambda ^j}{j!} \Big[r_kB_{\theta}(j,x-j+1) \sum_{i=1}^{k-1}\tau_i + \Gamma(j)^2 {\theta}^j \: {}_3\tilde{F}_2(j,j,j-x;j+1,j+1;\theta) \Big]}{r_k \sum_{j=1}^{x} \frac{\binom{x-1}{j-1} \lambda ^j}{j!} B_{\theta}(j,x-j+1)}.\nonumber\\
			\label{mean2} 
		\end{eqnarray}
		Recall that for the single-phase process, $\theta = \theta' = 1$, so
		\begin{eqnarray}
			\mathbb{E}(t) &=& \frac{ \psi(x+1)}{r_1} - \frac{\sum_{j=1}^{x} \frac{\lambda ^j }{j!} \psi(j)}{r_1 \left(e^{\lambda}-1 \right)}.
			\label{mean21}
		\end{eqnarray}
		
\item{\textbf{Variance}} \\
The variance is given by $\mathbb{E}(t^2) - \mathbb{E}(t)^2$, where
		\begin{eqnarray}
			\mathbb{E}(t^2)&=& \frac{r_k}{\sum_{j=1}^{x} \frac{\binom{x-1}{j-1} \lambda ^j}{j!} B_{\theta}(j,x-j+1)} \sum_{j=1}^{x} \frac{\binom{x-1}{j-1} \lambda ^j}{j!} \overbrace{ \int_{\sum_{i=1}^{k-1}\tau_i}^{\infty} t^2 e^{-jz} (1 - e^{-z})^{x-j} dt}^{(D)},\nonumber\\
			\label{var21c}
		\end{eqnarray}
where $z$ is given by \eqref{z_eqn}.

		Let $w = e^{-z}. $
		Then,
		\begin{eqnarray}
			D &\stackrel{\text{see \eqref{var1}}}{=}& \frac{\Big(\frac{\partial ^2}{\partial n^2} + 2\frac{\partial ^2}{\partial n \partial x} + \frac{\partial ^2}{\partial x^2} \Big) B_{\theta}(n,x-n+1)}{{r_k}^3 } + \frac{2\ln\theta'\Gamma(n)^2 {\theta}^n \: {}_3\tilde{F}_2(n,n,n-x;n+1,n+1;\theta)}{r_k^3 } + \nonumber\\
            && B_{\theta}(j,x-j+1) \left( \frac{\left(\sum_{i=1}^{k-1}\tau_i \right)^2 - \left(\sum_{i=1}^{k-1}\frac{r_i\tau_i}{r_k} \right)^2}{r_k} \right)
			\label{vard}
		\end{eqnarray}
where $\theta'$ is given by \eqref{thetap_eqn}.

		Plugging \eqref{vard} into \eqref{var21c}, we obtain			
		\begin{eqnarray}
	\mathbb{E}(t^2)&=& \frac{1}{r_k^2 \sum_{j=1}^{x} \frac{\binom{x-1}{j-1} \lambda ^j}{j!} B_{\theta}(j,x-j+1)} \sum_{j=1}^{x} \frac{\binom{x-1}{j-1} \lambda ^j}{j!} \Biggr[\Big(\frac{\partial ^2}{\partial n^2} + 2\frac{\partial ^2}{\partial n \partial x} + \frac{\partial ^2}{\partial x^2} \Big) B_{\theta}(n,x-n+1) + \nonumber\\
    &&2\ln\theta'\Gamma(n)^2 {\theta}^n \: {}_3\tilde{F}_2(n,n,n-x;n+1,n+1;\theta) + r_k^2B_{\theta}(j,x-j+1) \left(\left(\sum_{i=1}^{k-1}\tau_i \right)^2 - \left(\sum_{i=1}^{k-1}\frac{r_i\tau_i}{r_k} \right)^2 \right) \Biggl].\nonumber\\
    .
	\label{var2}
		\end{eqnarray}
For the single-phase process, the variance is given by
		\begin{eqnarray}
	\textbf{Var}(t) &=& \frac{ (e^{\lambda}-1)\sum_{j=1}^{x} \frac{\lambda^j}{j!}  \Big[\psi_1(j) + \psi(j)^2 \Big] - \Bigg( \sum_{j=1}^{x} \frac{\lambda^j}{j!}  \psi(j) \Bigg)^2}{\left(r_1 (e^{\lambda}-1) \right)^2} - \frac{\psi_1(x+1)}{r_1^2}.
	\label{var21}
		\end{eqnarray}
		
		\item \textbf{cdf} \\
		The cdf of the $Ct$ value is given by
		\begin{eqnarray}
			F(t|\lambda, \vec{r}, \vec{\tau}, x) &=& \frac{r_k \sum_{j=1}^{x} \frac{\binom{x-1}{j-1} \lambda ^j}{j!} \int_{\sum_{i=1}^{k-1}\tau_i}^{t} e^{-jz'} (1 - e^{-z'})^{x-j} ds}{\sum_{j=1}^{x} \frac{\binom{x-1}{j-1} \lambda ^j}{j!} B_{\theta}(j,x-j+1) }, \label{cdf2c}
		\end{eqnarray}
		where $z' = r_ks + \sum_{i=1}^{k-1}(r_i-r_k)\tau_i.$

Let $w = e^{-z'}$. Then,
	\begin{eqnarray}
	F(t|\lambda,\vec{r},\vec{\tau},x) &=& \frac{ \sum_{j=1}^{x} \frac{\binom{x-1}{j-1} \lambda ^j}{j!} \int_{e^{-z}}^{\theta} w^{j-1} (1 - w)^{x-j} dw}{\sum_{j=1}^{x} \frac{\binom{x-1}{j-1} \lambda ^j}{j!} B_{\theta}(j,x-j+1)} \nonumber\\
	&=& \frac{ \sum_{j=1}^{x} \frac{\binom{x-1}{j-1} \lambda ^j}{j!} \Big[ \int_{0}^{\theta} w^{j-1} (1 - w)^{x-j} dw - \int_{0}^{e^{-z}} w^{j-1} (1 - w)^{x-j} dw \Big]}{\sum_{j=1}^{x} \frac{\binom{x-1}{j-1} \lambda ^j}{j!} B_{\theta}(j,x-j+1)} \nonumber\\
	&=& \frac{ \sum_{j=1}^{x} \frac{\binom{x-1}{j-1} \lambda ^j}{j!} \Big[ B_{\theta}(j,x-j+1) - B_{e^{-z}}(j,x-j+1) \Big]}{\sum_{j=1}^{x} \frac{\binom{x-1}{j-1} \lambda ^j}{j!} B_{\theta}(j,x-j+1)} \nonumber\\
	&=& 1 - \frac{\sum_{j=1}^{x} \frac{\binom{x-1}{j-1} \lambda ^j}{j!} B_{e^{-z}}(j,x-j+1)}{\sum_{j=1}^{x} \frac{\binom{x-1}{j-1} \lambda ^j}{j!} B_{\theta}(j,x-j+1)},
	\label{cdf2}
\end{eqnarray}
where $\theta$ is given by \eqref{theta_eqn}.

For the single-phase process, using \eqref{simplification}, we simplify the cdf to obtain
	\begin{eqnarray}
	F(t|\lambda,r_1,x) &=& 1 - \frac{x \sum_{j=1}^{x} \frac{\binom{x-1}{j-1} \lambda ^j}{j!} B_{e^{-r_1t}}(j,x-j+1)}{e^{\lambda}-1}.
		\label{cdf21}
\end{eqnarray}

\item \textbf{Probability density of $\lambda$}\\
We conclude by deriving the probability density of $\lambda$, $P(\lambda|r_1,t,x)$, for the single-phase process. This density can be used to estimate $\lambda$ from measured $Ct$ values, and for calculating both the LoD and the LoQ of a PCR process. The steps described below can also be used to derive $P(\lambda|\vec{r},t,\vec{\tau})$, for a PCR process with an arbitrary number of phases.

By Bayes' Theorem, we have
\begin{eqnarray}
P(\lambda|r_1,t,x) &\propto & \frac{ \sum_{j=1}^{x} \frac{\binom{x-1}{j-1}}{j!} \left(\frac{\lambda w}{1-w} \right)^j}{e^{\lambda} - 1 },
\end{eqnarray} 
where $w = e^{-r_1t}$.

The normalizing constant is given by
\begin{eqnarray}
C &=& \int_{0}^{\infty}{ \frac{ \sum_{j=1}^{x} \frac{\binom{x-1}{j-1}}{j!} \left(\frac{\lambda w}{1-w} \right)^j}{e^{\lambda} - 1 } d \lambda}\nonumber\\
&=& \sum_{j=1}^{x} \frac{\binom{x-1}{j-1}}{j!} \left(\frac{w}{1-w} \right)^j \int_{0}^{\infty} \frac{\lambda^j}{e^{\lambda} - 1 } d \lambda \nonumber\\ 
&=& \sum_{j=1}^{x} \frac{\binom{x-1}{j-1}}{j!} \left(\frac{w}{1-w} \right)^j \Gamma(j+1) \zeta(j+1) \nonumber\\
&=& \sum_{j=1}^{x} \binom{x-1}{j-1} \left(\frac{w}{1-w} \right)^j \zeta(j+1),
\end{eqnarray} 
where $\zeta(j)$ is the Riemann zeta function.

Therefore, the probability density of $\lambda$ is given by
\begin{eqnarray}
P(\lambda|r_1,t,x) &=& \frac{\sum_{j=1}^{x} \frac{\binom{x-1}{j-1}}{j!} \left(\frac{\lambda w}{1-w} \right)^j}{(e^{\lambda}-1) \sum_{j=1}^{x} \binom{x-1}{j-1} \left(\frac{w}{1-w} \right)^j \zeta(j+1) } \nonumber\\
&=& \frac{\lambda w \; {}_1F_1(1-x,2,\frac{-\lambda w}{1-w})}{(e^{\lambda}-1)(1-w) \sum_{j=1}^{x} \binom{x-1}{j-1} \left(\frac{w}{1-w} \right)^j \zeta(j+1) }.
	\label{pdflam}
\end{eqnarray}
The probability that $\lambda$ takes values between $a$ and $b$ is given by
\begin{eqnarray}
P(a \leq \lambda \leq b \:| \:r_1,t,x) &=& \frac{ \sum_{j=1}^{x} \frac{\binom{x-1}{j-1}}{j!} \left(\frac{w}{1-w} \right)^j \int_{a}^{b} \frac{s^j}{e^{s}-1} ds}{\sum_{j=1}^{x} \binom{x-1}{j-1} \left(\frac{w}{1-w} \right)^j \zeta(j+1) } \nonumber\\
&=& \frac{ \sum_{j=1}^{x} \binom{x-1}{j-1} \left(\frac{w}{1-w} \right)^j \left[\zeta_b(j+1) - \zeta_a(j+1) \right] }{\sum_{j=1}^{x} \binom{x-1}{j-1} \left(\frac{w}{1-w} \right)^j \zeta(j+1) },
\label{pdflam2}
\end{eqnarray}
where $\zeta_{\lambda}(\cdot)$ is the incomplete Riemann zeta function.

It follows that the cumulative density function of $\lambda$ is given by
\begin{eqnarray}
F(\lambda \:| \:r_1,t,x) &=& \frac{ \sum_{j=1}^{x} \binom{x-1}{j-1} \left(\frac{w}{1-w} \right)^j \zeta_{\lambda}(j+1) }{\sum_{j=1}^{x} \binom{x-1}{j-1} \left(\frac{w}{1-w} \right)^j \zeta(j+1) }.
\label{cdflam}
\end{eqnarray}

As discussed earlier in relation to $P(n|\vec{r},t,\vec{\tau},x)$, Equation \eqref{pdflam} can be interpreted as follows: 
Suppose that a $Ct$ value $t$ is produced by a PCR process with expected number of input molecules $\lambda$. If we replace $\lambda$ by an estimate $\hat{\lambda}$, then \eqref{pdflam} gives the likelihood of $\hat{\lambda}$. For practical purposes (eg. to determine the LoQ), it is useful to calculate the probability that $\hat{\lambda}$ will be obtained as the estimate of $\lambda$ from any data $t$ that can be produced by a PCR process with expected number of input molecules $\lambda$. This probability is given by
\begin{eqnarray}
P\left(\hat{\lambda}|\lambda,r_1,x\right) &=& \int_{\sum_{i=1}^{k-1}\tau_i}^{\infty}P\left(\hat{\lambda},t|\lambda,r_1,x\right)dt\nonumber\\
&=& \int_{\sum_{i=1}^{k-1}\tau_i}^{\infty}P\left(\hat{\lambda}|\lambda,r_1,t,x\right)P\left(t|\lambda,r_1,x\right)dt\nonumber\\
&=& \int_{\sum_{i=1}^{k-1}\tau_i}^{\infty}P\left(\hat{\lambda}|r_1,t,x\right)P\left(t|\lambda,r_1,x\right)dt\nonumber\\
&=& \frac{r_1 x}{\left(e^{\lambda}-1 \right)\left(e^{\hat{\lambda}}-1 \right)} \int_{\sum_{i=1}^{k-1}\tau_i}^{\infty}  \frac{  (1-w)^{x} \Big(\sum_{j=1}^{x} \frac{\binom{x-1}{j-1}}{j!} \left(\frac{\lambda w}{1-w} \right)^j \Big)\Big(\sum_{j=1}^{x} \frac{\binom{x-1}{j-1}}{j!} \left(\frac{\hat{\lambda} w}{1-w} \right)^j \Big)}{ \sum_{j=1}^{x} \binom{x-1}{j-1} \left(\frac{w}{1-w} \right)^j \zeta(j+1) } dt \nonumber\\
&=& \frac{ x}{\left(e^{\lambda}-1 \right)\left(e^{\hat{\lambda}}-1 \right)} \int_{0}^{\theta}  \frac{  (1-w)^{x} \Big(\sum_{j=1}^{x} \frac{\binom{x-1}{j-1}}{j!} \left(\frac{\lambda w}{1-w} \right)^j \Big)\Big(\sum_{j=1}^{x} \frac{\binom{x-1}{j-1}}{j!} \left(\frac{\hat{\lambda} w}{1-w} \right)^j \Big)}{ w \sum_{j=1}^{x} \binom{x-1}{j-1} \left(\frac{w}{1-w} \right)^j \zeta(j+1) } dw,\nonumber\\
\label{like22}
\end{eqnarray}
where $\theta$ is given by \eqref{theta_eqn} and we have assumed that $\hat{\lambda}$ is conditionally independent of $\lambda$ given $t$.

It follows that the $t$-independent probability that $\hat{\lambda}$ will take values between $a$ and $b$ is given by
\begin{equation}
P(a \leq \hat{\lambda} \leq b \:| \:\lambda,r_1,x) = \int_{a}^{b} P\left(\hat{\lambda}|\lambda,r_1,x\right) d\hat{\lambda}.
\label{lamhat}
\end{equation}
	\end{itemize}

 \newpage

	\subsection{Supplementary Figures}\label{suppl}

\begin{figure}[!htb]
		\centering
	\includegraphics[width=\textwidth]{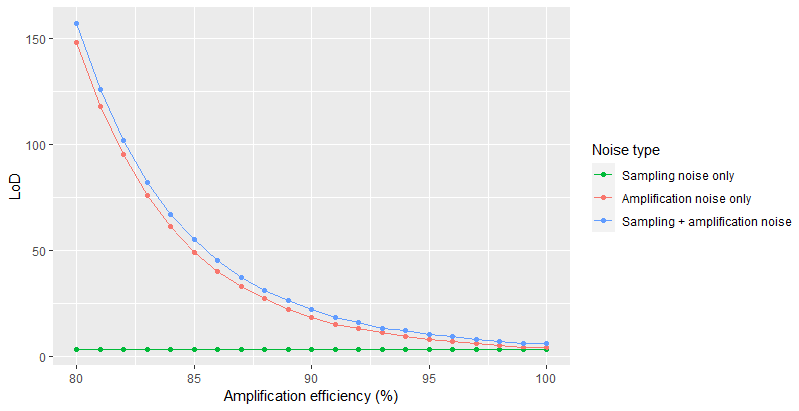}
	\caption{ \textbf{Limit of detection of the single-phase process}. The LoD was determined while accounting for either sampling noise alone (solid green line), amplification noise alone (solid red line), or both sampling noise and amplification noise (solid blue line). It was then plotted versus amplification efficiency, which is expressed on a base-2 scale as a percentage. The LoD based on sampling noise alone equals 3, whereas the LoD is highest when accounting for both types of noise. In the latter case, it ranges from 157, when the efficiency is only 80\%, to 6, when the efficiency is 100\%. The plot shows a strong dependence of the LoD on efficiency.}
	\label{fig:appdxfig2}
\end{figure}
	
\begin{figure}
		\centering
	\includegraphics[width=\textwidth]{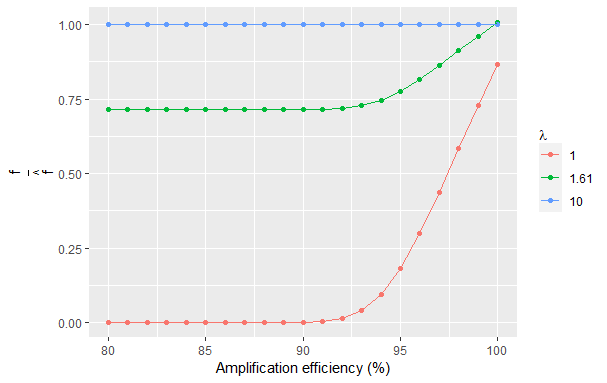}
	\caption{ \textbf{Ratio of expected versus estimated fraction of positive partitions in digital PCR}. For different expected numbers of input molecules $\lambda$, the ratio of the fraction of digital PCR partitions expected to test positive was calculated using Equation \eqref{cdf21t} and divided by the standard estimate based on the Poisson distribution (i.e. $1-e^{-\lambda}$). The result was plotted versus the amplification efficiency. While the efficiency used in calculations is always expressed on a base-$e$ scale, for ease of comprehension it was converted into a base-2 scale and displayed as a percentage.}
	\label{fig:appdxfig3}
\end{figure}


	\newpage
	\bibliography{references}
	
\end{document}